\documentclass[referee]{raa}            

\usepackage{graphicx,times}
\usepackage{natbib}
\usepackage{amssymb,amsmath}
\bibpunct{(}{)}{;}{a}{}{,}           
\usepackage{longtable}
\usepackage[pagebackref=true]{hyperref}
\usepackage{subcaption}
\usepackage[export]{adjustbox}
\usepackage{graphicx} 
\usepackage{xcolor}
\usepackage{lineno}

\begin{document}
   
   \title{Search for Periodic Radio Signals from Double Neutron Star System Companions Using the Fast Folding Algorithm} 
   
   \volnopage{Vol.0 (200x) No.0, 000--000}      
   \setcounter{page}{1}          

   \author{Wenze Li 
      \inst{1}
   \and Zhichen Pan
      \inst{2,3,4,5}
   \and Lei Qian
      \inst{2,3,4,5}
   \and Liyun Zhang
      \inst{1,6}
   \and Yujie Chen
      \inst{1}
    \and Dejiang Yin
      \inst{1}
    \and Baoda Li
      \inst{1}
    \and Yinfeng Dai
      \inst{7,8}
    \and Yaowei Li
      \inst{9}
    \and Dongyue Jiang
      \inst{1}
    \and Qiaoli Hao
      \inst{2}
    \and Menglin Huang
      \inst{2}
    \and Xingyi~Wang
      \inst{2}
    \and Xianghua Niu
      \inst{2}
    \and Minglei Guo
      \inst{2}
    \and Jinyou Song
      \inst{2}
    \and Shuangyuan~Chen
      \inst{2}
   }

   \institute{College of Physics, Guizhou University, Guiyang 550025, China; {\it liy\_zhang@hotmail.com}\\
        \and
             National Astronomical Observatories, Chinese Academy of Sciences,
             Beijing 100101, China; {\it panzc@nao.cas.cn; lqian@nao.cas.cn}\\
        \and
             Guizhou Radio Astronomical Observatory, Guizhou University,
             Guiyang 550025, China\\
        \and
             College of Astronomy and Space Sciences, University of Chinese Academy of Sciences,
             Beijing 100049, China\\
        \and
             Key Laboratory of Radio Astronomy, Chinese Academy of Sciences,
             Beijing 100101, China\\
        \and     
             International Centre of Supernovae, Yunnan Key Laboratory, Kunming 650216, China\\
        \and 
             School of Physics and Astronomy, Beijing Normal University, Beijing 100875, China\\
        \and 
            Department of Physics, Faculty of Arts and Sciences, Beijing Normal University, Zhuhai 519087, China\\
        \and 
            Department of Physics, School of Physics and Electronics, Hunan Normal University, Changsha 410081, China\\
         }  
   
   \date{Received~~2009 month day; accepted~~2009~~month day}

\abstract
{As most of the companions in the double neutron star systems should be normal pulsars,
the Fast Folding Algorithm (FFA), which is suitable for finding these long spin period pulsars,
was used to search their possible radio signals. A time domain resampling code PYSOLATOR was used to maximize the available data length by removing the orbital modulation. 
We collected and processed 272.2~hours observational data taken by the Five-hundred-meter Aperture Spherical radio Telescope (FAST) for the 13 double neutron star systems in its sky.
The signal-to-noise ratios of known pulsar signals are obviously improved by this search method,
including the detection of a faint pulsar signal which only saw by folding the data.
Unfortunately, no companion signals were found among all the 197962 candidates. Geodetic precession of the orbit could enhance detectability in future observations.
}
         
   \authorrunning{Li et al.}            
   \titlerunning{DNS system companion signals search with FFA}  
   \maketitle

\section{Introduction}
\label{sect:intro}

A double neutron star (DNS) system consists of two neutron stars, 
and normally at least one of which can be observed as a pulsar. 
The discovery of the first double neutron star system, 
B1913+16 \citep{1975ApJ...195L..51H}, 
indirectly confirmed the existence of gravitational waves \citep{1982ApJ...253..908T} 
and provides an ideal laboratory for testing gravitational theories. 
Among hundreds of binary pulsar systems, 
24 have been confirmed as DNS systems and with 11 more being candidates \footnote{\url{https://www3.mpifr-bonn.mpg.de/staff/pfreire/NS_masses.html}}.
If the periodic signals from both neutron stars can be observed,
such a DNS was called a double pulsar.

The J0737-3039A/B binary system is the first and only double pulsar till now \citep{2003Natur.426..531B, 2004HEAD....8.1101L}. 
It is with a short orbital period $P_{b}$ = 2.4 hours) and moderate eccentricity $e$ = 0.088). 
The pulsars A and B have spin periods of 22.7~ms and 2.7~s, respectively. 
It is one of the best probes for validating general relativity and other theories of gravity in the strong regime field \citep{2008ARA&A..46..541K}. 
The five Keplerian parameters (Orbital Period $P_{b}$, Projected Semi-major Axis $x$, Eccentricity $e$, Longitude of Periastron $\omega_{0}$, Epoch of Periastron $T_{0}$) of each pulsar
and seven post-Keplerian parameters (Periastron Advance $\dot{\omega}$, Einstein Delay $\gamma$, Shapiro Delay Range $r$, Shapiro Delay Shape $s$, Orbital Decay $\dot{P}_{b}$, Relativistic Orbital Deformation $\delta_{\theta}$, Spin Precession of $B$ $\Omega_{\mathrm{B}}^{\text {spin }}$)
in this system were detected \citep{2021PhRvX..11d1050K}. 
The orbital decay rate $\dot{P}_b$ agrees with the General Relativity's gravitational radiation prediction within $1.3\times10^{-4}$, a precision $\sim25$ times that of the Hulse--Taylor pulsar \citep{2021PhRvX..11d1050K}.
In addition, double pulsar systems have unique significance for binary star evolution \citep{2006csxs.book..623T,2017ApJ...846..170T}, 
gravitational wave detection \citep{2017PhRvL.119p1101A}, 
neutron star internal structure and equation of state \citep{2016PhR...621..127L}, 
and extreme-environment plasma and magnetosphere \citep{2008Sci...321..104B}. 

Since the discovery of the DNS system, 
detecting more such systems has remained a long-standing ambition in pulsar surveys. 
The Five-hundred-meter Aperture Spherical radio Telescope (FAST, \citealt{2011IJMPD..20..989N}) is the largest single-dish radio telescope, 
offering exceptional sensitivity for detecting such objects. 
It has discovered over 1000 pulsars, 
including 5 DNSs among approximately 150 new binary systems: 
J1953+1847D \citep{2025ApJS..279...51L} in globular cluster M71, 
J1901+0658 \citep{2024MNRAS.530.1506S}, 
J0528+3529, J1844-0128\citep{2025RAA....25a4003W}, 
and J2150+3427 \citep{2023ApJ...958L..17W} in Galactic field. 
It is estimated that more than 100 DNS systems should be discovered by the FAST and the Square Kilometre Array (SKA) \citep{2015aska.confE..40K}. 
The second double pulsar system may be detectable with FAST using the FFA.

Most pulsars in known DNS systems are recycled, while their companions are considered to be normal, long-period pulsars according to pulsar evolution theory \citep[e.g.][]{2017ApJ...846..170T}.
The Fast Folding Algorithm (FFA) is a phase-coherent search method that detects periodic signals by directly folding time series \citep{1969IEEEP..57..724S}. 
Compared to the search pipeline based one Fast Fourier Transform (FFT),
FFA is more sensitive to pulse signals with long periods and low duty cycles \citep{2020MNRAS.497.4654M},
and was successfully used to find new pulsars in FAST data (e.g., the slowest globular cluster pulsar M15L, P $\sim$ 3961 ms, \citealt{2024ApJ...974L..23W}). 
The FFA also has an exceptionally high detection rate ($93.5\%$) and demonstrates superior sensitivity to weak signals even millisecond pulsars \citep{2025ApJ...991...38L}.
Therefore, FFA should be the suitable search algorithm to search for the companion periodic signals of DNS systems.

In this study, 
we employed the FFA to search for companion signals in DNS systems in the FAST sky.
The structure of this paper is as follows:
Section \ref{sect:data} describes the data and reduction process,
Section \ref{sect:results} presents the results,
Section \ref{sect:discussion} is the discussion,
and Section \ref{sect:conclusion} is the conclusion.

\section{Data and Data Reduction}
\label{sect:data}

\subsection{Data}

\begin{figure}[htbp]
  \centering
  \includegraphics[width=1.0\textwidth]{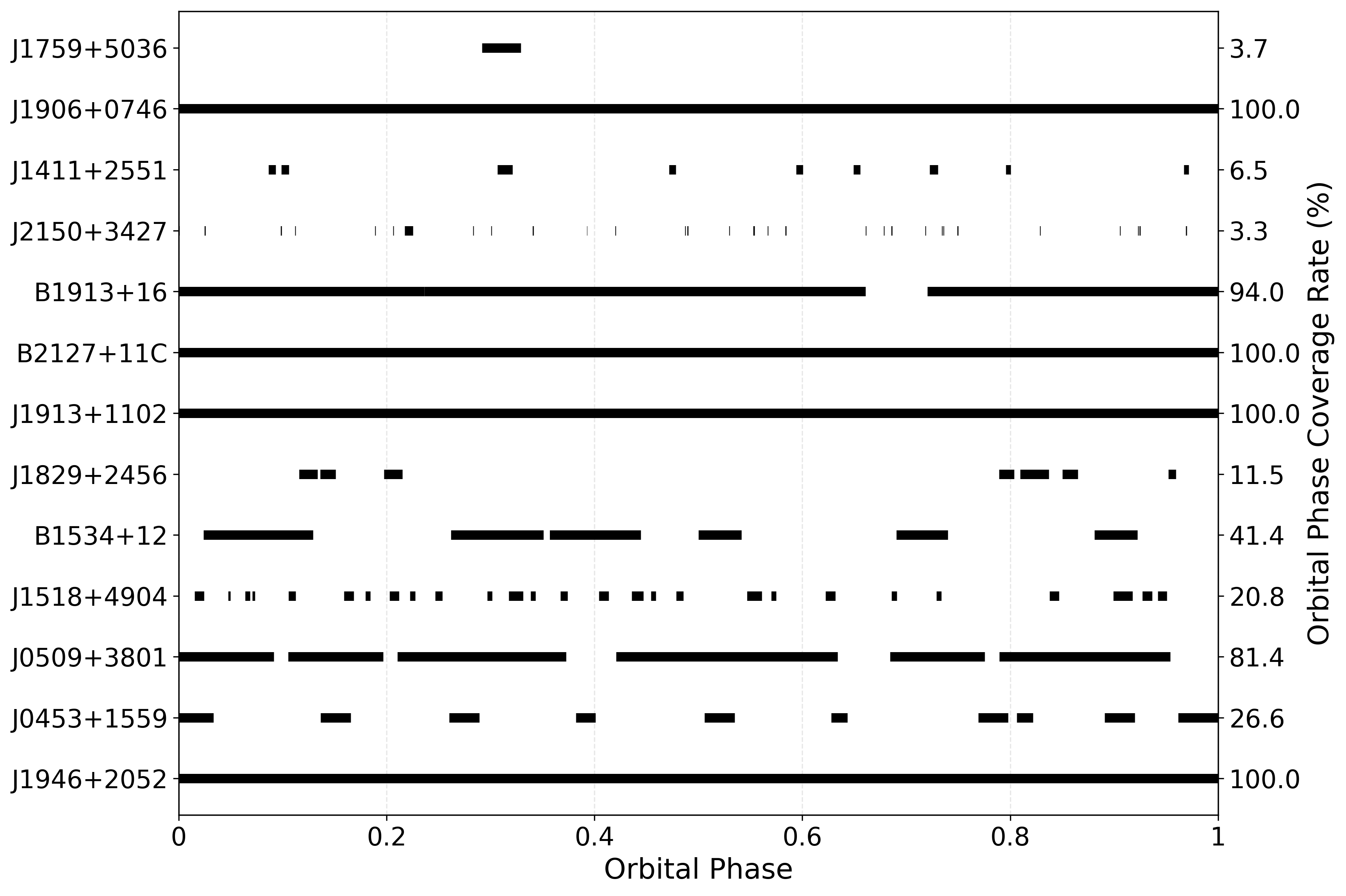} 
  \caption{The orbital phase coverage.}   
  \label{fig:phase_coverage}       
\end{figure}

There are 12 DNSs and one DNS candidate in the FAST sky.
The information of these 13 sources are listed in Table \ref{tab:table basic information}.
The observations were done with the FAST 19-beam receiver, 
which covers a frequency range of 1.05 to 1.45~GHz. 
The sampling time is 49.152~$\mu$s, 
and the system temperature is $\rm \sim$24~K \citep{2019SCPMA..6259502J}. 
The data was channelized into 4096 channels and corresponding to 0.122~MHz channel width, 
which packetized and stored in search-mode \textsc{PSRFITS} format. 
We collected as many data as we can from FAST data archive (see details in the FAST observation catalog \footnote{\url{https://fast.bao.ac.cn/}}). 
The data spanning, the number of observations and total observing time of each source can be found in Table \ref{tab:table observation data}.
The orbital phase coverage rate of each system is shown in Fig. \ref{fig:phase_coverage}. 
No eclipsing events are observed in any of our data.

\begin{table}[!ht]
    \centering
    \caption{Information of the 13 DNS systems in the FAST sky.}
    \label{tab:table basic information}
\begin{tabular}{ccccccccc}
\hline
Name & DM & $P$ &$P_{\rm b}$ & $\tau_{\rm age}$ & $e$ & $M_{\rm TOT}$ & $M_{\rm c}$ & $i$  \\
  & $(\text{pc cm}^{-3})$ & (s) &(days) & (Myr)  &  & (\( M_\odot \)) & (\( M_\odot \)) & ($^\circ$)  \\
\hline
J0453+1559$^{1}$ & 30.30 & 0.04578 & 4.072 & $\sim 4100$  & 0.11251847(8) & 2.734(4) & 1.174(4) & $75.7_{-0.8}^{+0.7}$ \\
J0509+3801$^{2}$ & 69.08 & 0.07654 & 0.380 & $\sim 200$  & 0.586409(3)& 2.81071(14) & 1.412(6) & $\star$ \\
J1411+2551$^{3}$ & 12.37 & 0.06245 & 2.616 & $\gtrsim 9100$  & 0.1699308(4)&2.538(22)&$>$ 0.92&$\star$  \\
J1518+4904$^{4}$ & 11.61 & 0.04093 & 8.634 & $\gtrsim 16000$  & 0.249484383(9)&2.7186(7)&$1.248(^{+35}_{-29})$& $49.6_{-1.8}^{+1.6}$ \\
B1534+12$^{5}$   & 11.62 & 0.03790 & 0.421 & $\sim 200$  & 0.27367740(4)&2.678463(4)&1.3455(2)& 77.7 ± 0.9  \\
J1759+5036$^{2}$ & 7.78  & 0.17602 & 2.043 & $\sim 50000$  & 0.30827(12)&2.679(12)&$>$ 0.83& $\star$ \\
J1829+2456$^{6}$ & 13.70  & 0.04101 & 1.176 & $\sim 12400$  & 0.1391435(3)&2.60551(38)&1.299(7)&75.8 ± 0.7 \\
J1906+0746$^{7}$ & 217.75 & 0.14407 & 0.166 & $\sim 0.112$  & 0.0852996(6)&2.6134(3)&1.322(11)&43.7 ± 0.5 
 \\
J1913+1102$^{8}$ & 338.96 & 0.02729 & 0.206 & $\sim 2800$  & 0.089531(2)&2.8887(6)&1.27(3)&55.3 \\
B1913+16$^{9}$  & 168.77 & 0.05903 & 0.323 & $\sim 100$  & 0.6171340(4)&2.828378(7)&1.390(1)&47.9 ± 0.1\\
J1946+2052$^{10}$ & 93.97 & 0.01696 & 0.078 & $\sim 290$  & 0.0638180(6)&2.54(3)&$>$ 1.2&$\star$ \\
B2127+11C$^{11}$ & 67.13 & 0.03053 & 0.335 & $\sim 97$  & 0.681395(2)&2.71279(13)&1.354(10)& 65 ± 2 
 \\
J2150+3427$^{12}$ & 55.48 & 0.65428 & 10.592 & $\sim 2880$  & 0.601494(2)&2.59(13)&$>$ 0.98&$\star$ \\
\hline
\end{tabular}
\begin{flushleft}  
\textbf{Note:} The column, from left to right, list the DNS system name, 
dispersion measure,
spin period,
orbital period, 
characteristic age
orbit eccentricity, 
the system mass, 
companion mass, 
and orbital inclination angle for each system. 
References: 1 \citep{2015ApJ...812..143M}, 2 \citep{2024ApJ...962..167M}, 3 \citep{2017ApJ...851L..29M}, 4 \citep{2024ApJ...966...26T}, 5 \citep{2014ApJ...787...82F}, 6 \citep{2021MNRAS.500.4620H}, 7 \citep{2015ApJ...798..118V}, 8 \citep{2020Natur.583..211F}, 9 \citep{2016ApJ...829...55W}, 10 \citep{2018ApJ...854L..22S}, 11 \citep{2006ApJ...644L.113J,2024ApJ...974L..23W}, 12 \citep{2023ApJ...958L..17W}.
\end{flushleft}
\end{table}

\begin{table}[!ht]
    \centering
    \caption{Observation summaries of DNS systems}
    \label{tab:table observation data}
    \begin{tabular}{cccccc}
    \hline 
        Name & $\rm Observation_{\rm start}$  & $\rm Observation_{\rm end}$  & Times & Total time  \\ 
        & yyyy--mm--dd & yyyy--mm--dd &  &  hr &                    \\
    \hline     
        J0453+1559 & 2019-10-16 & 2021-10-23 & 11 & 30.0 &  \\ 
        J0509+3801 & 2020-05-04 & 2023-05-21 & 16 & 13.0 &  \\ 
        J1411+2551 & 2020-08-26 & 2023-04-11 & 8  & 4.1  &  \\ 
        J1518+4904 & 2019-10-22 & 2023-01-01 & 32 & 46.5 &  \\ 
        B1534+12   & 2020-08-26 & 2023-05-21 & 21 & 10.1 &  \\ 
        J1759+5036 & 2024-12-21 & 2024-12-21 & 1  & 1.8  &  \\ 
        J1829+2456 & 2019-10-17 & 2023-07-25 & 8  & 3.3  &  \\ 
        J1906+0746 & 2019-10-18 & 2023-04-15 & 23 & 39.5 &  \\ 
        J1913+1102 & 2020-09-11 & 2023-04-21 & 23 & 18.5 &  \\ 
        B1913+16   & 2019-10-19 & 2023-04-20 & 26 & 20.3 &  \\ 
        J1946+2052 & 2020-10-27 & 2024-11-01 & 26 & 36.1 &  \\ 
        B2127+11C  & 2019-11-09 & 2023-05-21 & 20 & 40.6 &  \\ 
        J2150+3427 & 2022-05-03 & 2023-04-24 & 30 & 8.4  &  \\ 
    \hline
    \end{tabular}
    \begin{flushleft}  
    \textbf{Note:}  
       Columns 2 and 3 list the start and end dates of observations, respectively.
       Column 4 is for the total number of observations.
       Column 5 is for the cumulative observing time.  
    \end{flushleft}
\end{table}

\subsection{Data Reduction}

The \textsc{PRESTO} \footnote{\url{https://github.com/scottransom/presto}} \citep{2001PhDT.......123R,2002AJ....124.1788R,2003ApJ...589..911R} was used to pre-process the data.
The routine {\tt rfifind} was used to mask radio frequency interference.
The length of the data used to identify frequency domain interferences (-time option) was set to be 2 s.
The {\tt prepsubband} command was used to dedisperse the data around the pulsar’s dispersion measure (DM).
We generated eleven dedispersed time series for each observation, using DM values spaced by 0.1~pc~cm$^{-3}$ around the known pulsar's DM to ensure optimal coverage for potential candidate detections.
To avoid the influence of the pulsar's signal and its harmonics on the candidate identification in the search results, 
a birdie file was created to mask the pulsar’s signal.
The birdie masking is applied conservatively. The known pulsars are usually recycled pulsars (except for PSRJ1906+0746), while their companions are normal pulsars with longer periods than the known pulsars. We only mask the known pulsars and their low-order harmonics, if the companion signal could lie close to or coincide with a harmonic of the known pulsar’s signal, which do not mask the companion’s signal completely.

The \textsc{PYSOLATOR} \footnote{\url{https://alex88ridolfi.altervista.org/pagine/pulsar_software_PYSOLATOR.html}} \citep{2020ascl.soft03012R} was used to correct the pulsar’s spin period variations induced by the orbital motion.
It removes the orbital modulation by subtracting the predicted Roemer delay and resampling the time series, 
and resampling the data to the binary barycenter.
It need the ephemeris file in which the total mass of the system (MTOT), 
companion mass (M2), 
and the $\sin$ value of the orbital inclination angle (SINI) should be included.
In a double neutron star system, the projected semi-major axis of the companion pulsar, $A2$, generally differs from that of the known pulsar, $A1$. The two are related through the mass ratio $q = M1/M2=A2/A1$ ($M1$ and $M2$ denote the masses of the known pulsar and its companion, respectively.), such that
$A2 = q A1$.
For 5 DNS systems among all the 13 in the FAST sky, 
the precise masses of the companion star and the orbital inclination angles have not been measured (see Table  \ref{tab:table basic information}). 
Therefore, the canonical pulsar mass of 1.35~\( M_\odot \) in DNS systems \citep{2013ApJ...778...66K} was used as the companion mass, 
and the corresponding orbital inclination angle for such systems was adopted.

The \textsc{RIPTIDE} \footnote{\url{https://github.com/v-morello/riptide}} \citep{2020MNRAS.497.4654M} package was used to perform FFA pulsar searches via its {\tt rffa} routine.
The range of folding periods was set to be 0.001 -- 60 seconds.
The candidates with the SNR $\geq$ 7 were selected for further analysis.
The command {\tt prepfold} from \textsc{PRESTO} was used to fold candidates. 

Due to the relativistic spin precession, 
the interpulse of PSR J1946+2052 changed from a single-peak to a double-peak shape from 2018 to 2021 \citep{2024ApJ...966...46M}. 
Therefore, the timing solutions before 2018 are not applicable to the current observational data to remove orbital modulation.
We performed a timing analysis for J1946+2052 with FAST observations. 
The \texttt{prepfold} routine was used to fold the data with parameters from the parfile of J1946+2052 \citep{2018ApJ...854L..22S}. 
The standard profile was from the Gaussian component fitting of the pulse profile from one of the 26 observational data of this pulsar with \texttt{pygaussfit.py}.
The Time-of-arrivals (TOAs) were obtained using \texttt{get\_TOA.py} routine from \textsc{PRESTO}. 
For the timing analysis, the \texttt{APTB} based on \textsc{PINT} \citep{2024ApJ...971..150S} software package is used to obtain a phase-connected timing solution.

\section{Results}
\label{sect:results}

\subsection{The Timing Solution for J1946+2052}

\begin{figure*}[htbp]
  \centering
  \includegraphics[width=1.0\textwidth]{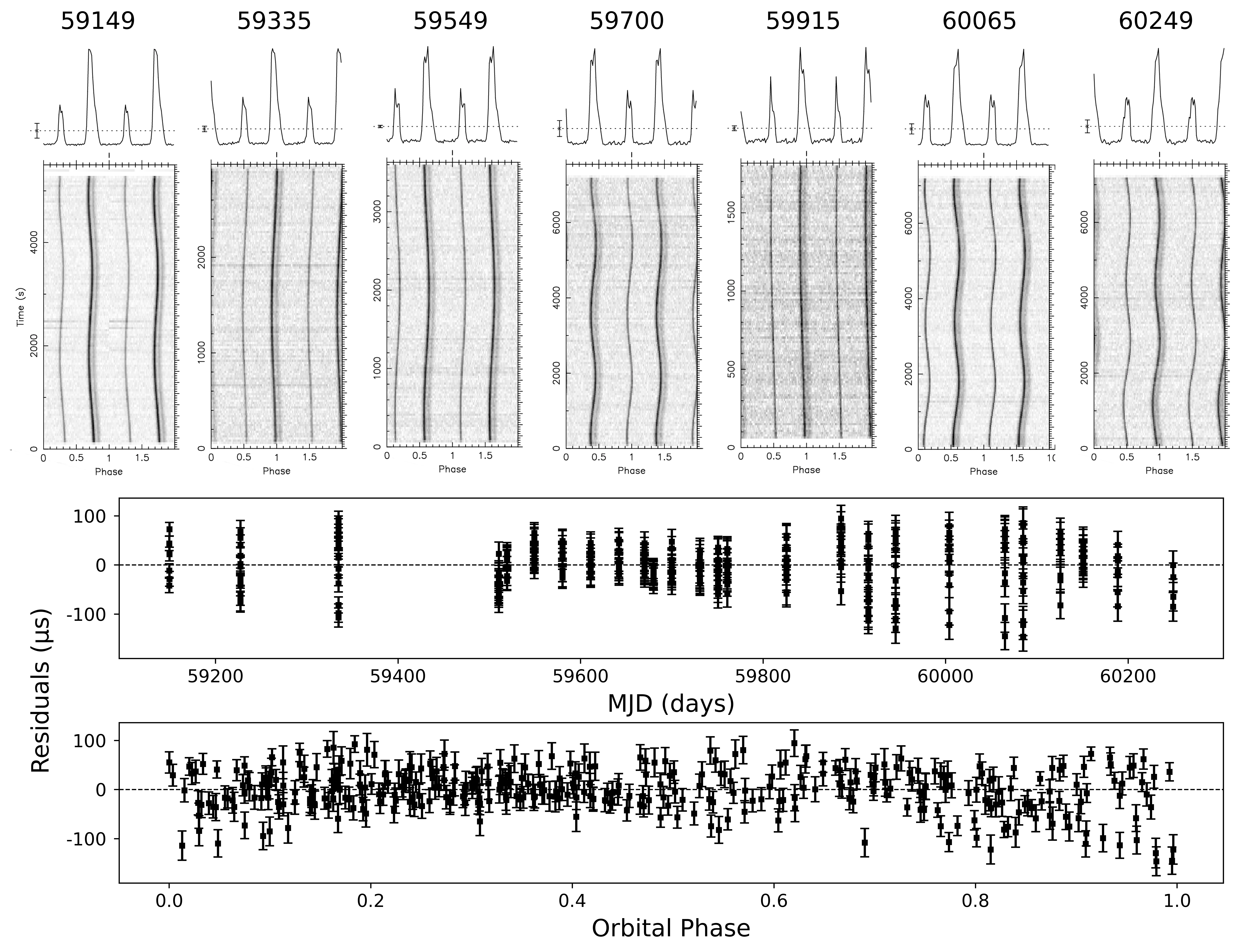} 
  \caption{The pulse profile (upper subplot) and the timing residual as a function of MJD and orbital phase from the best-fit timing model (lower subplot) with PSR J1946+2052.}   
  \label{fig:timing_change_pulse_profiles}       
\end{figure*}

We used the 26 observational data from FAST.
The time span is 1100 days, from 59149 to 60249. 
Our timing results and the previous one \citep{2018ApJ...854L..22S} are presented in Table \ref{tab:J1946_compare}, 
while the timing residuals are showed in Fig. \ref{fig:timing_change_pulse_profiles}.  
Using these orbital parameters, we obtain the mass function of \citep[e.g.][]{2016ARA&A..54..401O}
\begin{equation}
    f\left(m_{\mathrm{p}}, m_{\mathrm{c}}\right)=\frac{\left(m_{\mathrm{c}} \sin i\right)^{3}}{\left(m_{\mathrm{p}}+m_{\mathrm{c}}\right)^{2}}=\frac{4 \pi^{2} x^{3}}{P_{\mathrm{b}}^{2} T_{\odot}}=0.268137(2) \mathrm{M}_{\odot},
\label{eq:1}
\end{equation}
where $T_{\odot}=G M_{\odot}c^{-3}=4.925490947\ \mu$s, 
$i$ is the inclination angle of the orbit, 
and $m_{\rm p}$ and $m_{\rm c}$ are the masses of the pulsar and its companion, respectively. 
For pulsars in compact binary systems, 
the measurement of PK parameters can further constrain the masses of the pulsar and the companion. 
We measured periastron advance $\dot{\omega}$ only, which is 25.8696(1)~deg~yr${}^{-1}$. 
Assuming the periastron advance is caused by the general relativistic effect alone, 
the rate of periastron advance should be \citep[e.g.][]{2003LRR.....6....5S}
\begin{equation}
    \dot{\omega}=3T_{\odot}^{2/3}\left(\frac{P_{\mathrm{b}}}{2\pi}\right)^{-5/3}\frac{\left(m_{\mathrm{p}}+m_{\mathrm{c}}\right)^{2/3}}{1-e^{2}}.
    \label{eq:2}
\end{equation}
Thus, the total mass ($m_{\rm p}+m_{\rm c}$) was calculated as $2.54(3)\ M_{\odot}$, 
consistent with previous result \citep{2018ApJ...854L..22S} and the newly high-precision timing result \citep{2025arXiv251012506M}.

\begin{table*}[!ht]
\centering
\caption{Comparison of timing parameters for PSR~J1946+2052 between K.~Stovall~et~al.~(2018) and this work solution}
\label{tab:J1946_compare}
\begin{tabular}{lcc}
\hline
\hline
           & \cite{2018ApJ...854L..22S} & This Work \\
\hline
Right Ascension, $\alpha$ (J2000) & 19:46:14.130(6) & 19:46:14.13503(9) \\
Declination, $\delta$ (J2000) & +20:52:24.64(9) & +20:52:24.8999(3) \\
Spin Frequency, $f$ (s$^{-1}$) & 58.9616546384(5) & 58.961654623404(8) \\
1st Frequency Derivative, $\dot{f}$ (s$^{-2}$) & $-3.2(6)\times10^{-15}$ & $-3.7055(1)\times10^{-15}$ \\
2nd Frequency Derivative, $\ddot{f}$ (s$^{-3}$) & -- & $-1.2(8)\times10^{-24}$ \\
Reference Epoch (MJD) & 57989.0 & 57989.0 \\
Dispersion Measure, DM (pc cm$^{-3}$) & 93.965(3) & 93.965 \\
Solar System Ephemeris & DE436 & DE421 \\
Terrestrial Time Standard & TT(BIPM) & TT(BIPM2023) \\
Time Units & TDB & TDB \\
Span of Timing Data (MJD) & 57953–58024 & 59149–60249 \\
Residuals RMS ($\mu$s) & 95.04 & 42.53 \\
Number of TOAs & -- & 364 \\
\hline
\multicolumn{3}{c}{Binary Parameters} \\
\hline
Binary Model & DD & DD \\
Orbital Period, $P_b$ (days) & 0.07848804(1) & 0.0784880558834(4) \\
Projected Semi-major Axis, $x_{\rm p}$ (lt-s) & 1.154319(5) & 1.1544228(4) \\
Eccentricity, $e$ & 0.063848(9) & 0.0638180(6) \\
Longitude of Periastron, $\omega$ (deg) & 132.88(1) & 132.527(4) \\
Epoch of Periastron, $T_0$ (MJD) & 57989.002943(3) & 57989.0028971(9) \\
Periastron Advance, $\dot{\omega}$ (deg\,yr$^{-1}$) & 25.6(3) & 25.8696(1) \\
Mass Function, $f_{\rm mass}$ ($M_\odot$) & 0.268184(12) & 0.268137(2) \\
Total System Mass, $M_{\rm Total}$ ($M_\odot$) & 2.50(4) & 2.54(3) \\
\hline
\end{tabular}
\end{table*}

\subsection{The Search Results}

\begin{figure*}[htbp]
  \centering
  \includegraphics[width=1.0\textwidth]{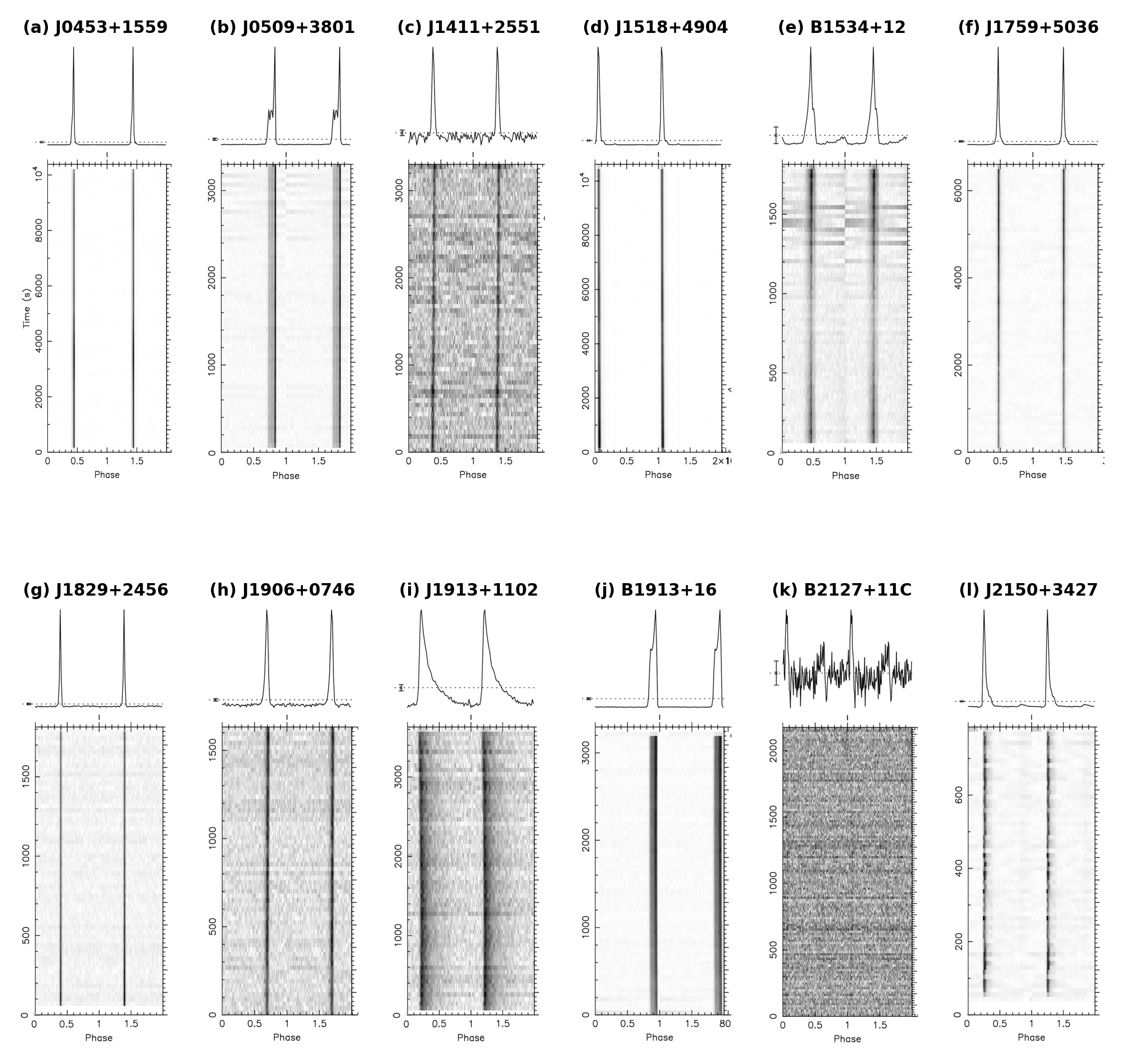} 
  \caption{The pulse profiles of the primary stars in 12 DNS systems, the results of FFA search after the orbital effects were removed.}
  \label{fig:phase_profile}       
\end{figure*}

We successfully removed the orbital effects of all the 13 DNS systems in FAST sky using \textsc{PYSOLATOR}. 
Fig. \ref{fig:phase_profile} shows the signals of the 12 pulsars in the DNS systems, 
after the orbital effects were removed.
For J1946+2052, the pulsar period in every observation still varies slightly due to the errors from the timing solution.
As the period variation introduces phase drift, it leads to pulse broadening and has an unknown impact on the sensitivity of the FFA-based search.
Some folding results from its resampled data can be found in Fig.\ref{fig:timing_change_pulse_profiles} as examples. We also aim to achieve more precise timing solutions \citep[such as the result of][]{2025A&A...704A.153M}, and the orbital effects can be further minimized or fully removed in future analyses.

We searched for the companion periodic radio signals of the 13 DNS systems, with a total of 245 observational data and a duration of 272.2~hours (see Table \ref{tab:table observation data}). 
As a result, 197962 candidates were found from all the data.
Every candidate was checked carefully and none was confirmed as the companion signals.

\clearpage

\section{Discussion}
\label{sect:discussion}

\subsection{The Search Sensitivities} 

\begin{figure*}[htbp]
  \centering
  \includegraphics[width=0.9\textwidth]{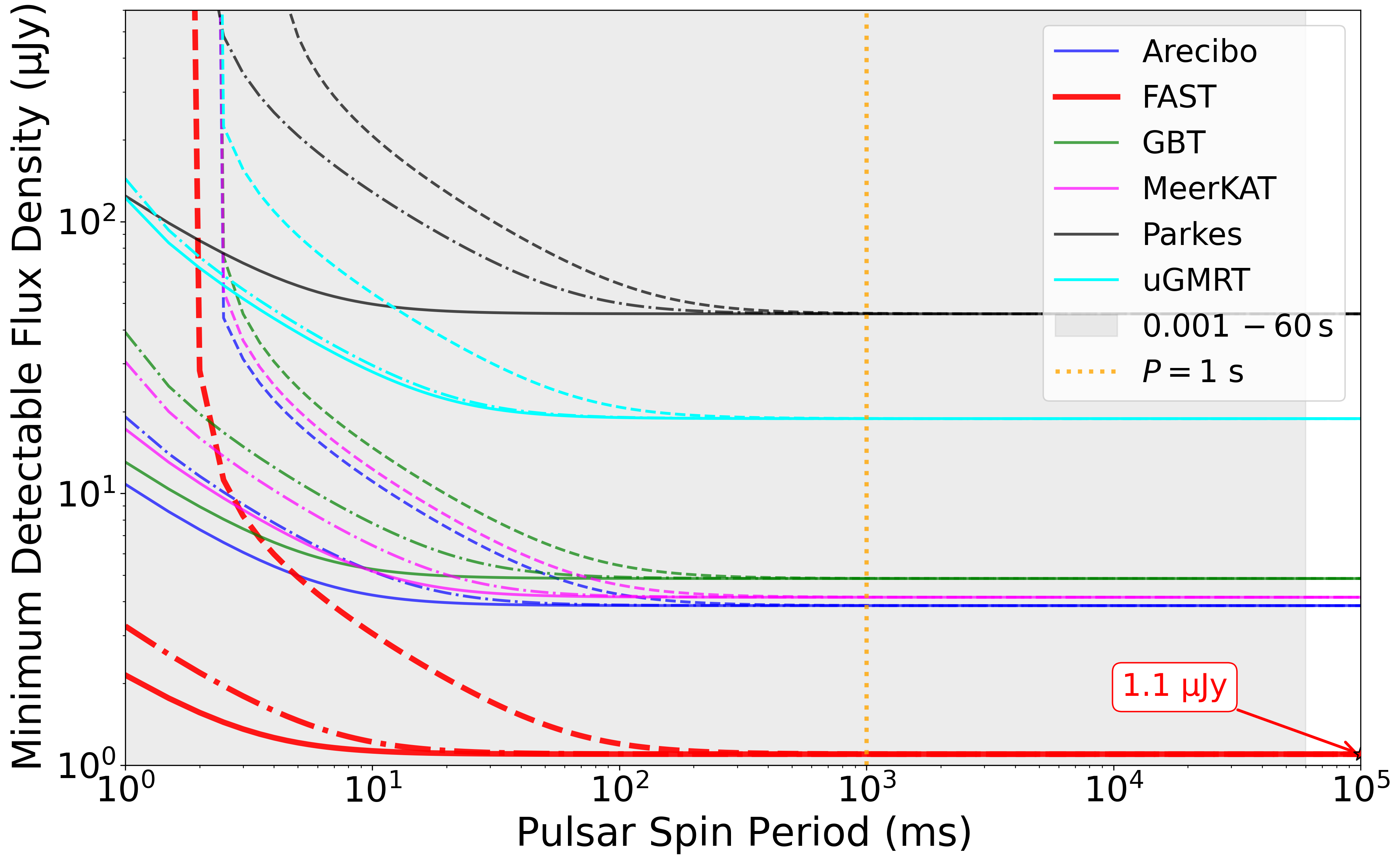} 
  \caption{Survey sensitivity as a function of pulsar period and DM
  when the observation time is 2~hours for all the telescopes, 
  assuming an intrinsic pulse width of 3 $\%$.
  The solid line, dash-dot line, and dash line in every colour represent the DM values of 
  10, 200, and 350~$\text{pc cm}^{-3}$, respectively, 
  as these three numbers are the lowest, median, and highest DM values among all the selected 13 pulsars in our study.
  The shaded area represents the searched pulsar period range.}
  \label{fig:fast}       
\end{figure*}

The search sensitivities were estimated using the radiometer equation \citep[e.g.][]{1985ApJ...294L..25D}
\begin{equation}
S_{\mathrm{min}}=\beta \frac{\left(S / N_{\mathrm{min}}\right) T_{\mathrm{sys}}}{G \sqrt{n_{\mathrm{p}} T_{\mathrm{obs}} \Delta f}} \sqrt{\frac{W_{\mathrm{obs}}}{P-W_{\mathrm{obs}}}},
\label{eq:sensitivities}
\end{equation}
in which the $\beta$ is the sampling efficiency (here we use 1 for the 8-bit recording system), 
the $(S/N_{\text{min}})$ is the minimum signal-to-noise ratio (being 7 in our search),
the number of polarizations $(n_{\text{p}})$ is 2, 
the $T_{\mathrm{obs}}$ is the integration time in seconds, 
the $W_{\mathrm{obs}}$ is the width of the pulsar profile, 
the $P$ is the spin period of pulsar.
For FAST, 
the system temperature $(T_{\text{sys}}$) is 24~K,
the antenna gain, $G$, is $16\,\mathrm{K\,Jy^{-1}}$, 
the $\Delta f$ is the bandwidth in the unit of $\mathrm{MHz}$, here is 400~MHz \citep{2019SCPMA..6259502J}.
The parameters of some other radio telescopes are listed in Table \ref{tab:table Telescope parameters}.
Assuming the $W_{\mathrm{obs}}$ is 3\% for long period pulsars, 
the sensitivity varying with different DM values and spin periods was present in Fig. \ref{fig:fast}.
As the orbital effects removed, 
the observing time (here we used is 2~hour) was used instead of considering the orbital effects during pulsar search.

FAST no doubt exhibits the highest sensitivity among all the telescopes.
When the DM value is 350~pc~cm$^{-3}$,
the sensitivity decreased a lot when the spin period is 2~ms or shorter.
Thus, excluding the situation that the companion is a very fast millisecond pulsar, 
our search reached a higher sensitivity.

On the other hand, the single pulse search will break such sensitivity estimation,
while the aperture is the most important parameter.
Geodetic precession of the orbit may enhance detectability, which should be taken into account in the future.

\begin{table*}[htbp]
\centering
\caption{Telescope parameters used in sensitivity calculations}
\label{tab:table Telescope parameters}
\begin{tabular}{lccccccc}
\hline
Telescope & $\beta$ & $T_{\rm sys}$ (K) & $G$ (K Jy$^{-1}$) & $\Delta f$ (MHz) & $\nu_{\rm center}$ (GHz) & Channels & Frequency range (GHz) \\
\hline
FAST$^{1}$     & 1.0  & 24.0   & 16.0  & 400 & 1.25  & 4096 & 1.05--1.45 \\
Arecibo$^{2}$  & 1.2  & 40.0   & 10.5  & 300 & 1.40  & 256 (512) & 1.15--1.55 \\
MeerKAT$^{3}$  & 1.1  & 26.0   & 2.8   & 650 & 1.284 & 1024 & 0.96--1.60 \\
GBT$^{4}$      & 1.05 & 22.8   & 2.0   & 650 & 1.50  & 512  & 1.18--1.83 \\
uGMRT$^{5}$    & 1.0  & 102.5  & 4.2   & 180 & 0.65  & 4096 & 0.56--0.74 \\
Parkes$^{6}$   & 1.5  & 35.0   & 0.7   & 288 & 1.374 & 96   & 1.23--1.52 \\
\hline
\end{tabular}
\begin{flushleft}  
\textbf{Note:} Column 2 lists the sampling efficiency that accounts for various sensitivity losses due to signal processing and digitization. Columns 3 and 4 list the system temperature and antenna gain, and Columns 5 and 6 list the bandwidth and central observing frequency. Column 7 lists the number of channels, Column 8 lists the low- and high-frequency edges of the bandwidth. To optimize channel selection, we set channels to be 256 if the DM value is $<$ 100~$\text{pc cm}^{-3}$. For higher DM values ($>$ 100~$\text{pc cm}^{-3}$), we set channels to be 512, ensuring sufficient resolution. References: 1 \citep{2019SCPMA..6259502J}, 2 \citep{2007ApJ...670..363H}, 3 \citep{2021MNRAS.504.1407R}, 4 \citep{2022ApJ...941...22M}, 5 \citep{2022A&A...664A..54G}, 6 \citep{2000ApJ...535..975C}.
\end{flushleft}
\end{table*}

\clearpage

\subsection{FFA Plus \textsc{PYSOLATOR}}

The FFA shows significant advantages in searching for long-period pulsars with long spin periods~($P \sim 1 \rm s$) compared to the traditional FFT-based method \citep{2017MNRAS.468.1994C}. 
This advantage arises from its time-domain folding technique, 
which avoids the sensitivity loss caused by red noise in FFT-based searches. 
Furthermore, FFA produces fully phase-coherent folded pulse profiles, 
while FFT methods can only incoherently sum a limited number of harmonics (typically 16–32), 
leaving part of power unused. 
In terms of computational efficiency, 
FFA scales as $O(N \log_{2}(N/P_{0}))$ \citep{2017MNRAS.468.1994C}, 
where $N$ is the number of data samples and $P_{0}$ is the base trial period. 
As $P_{0}$ increases, FFA becomes increasingly efficient, 
making it well-suited for systematic searches of isolated, long-period pulsars.  

Despite these advantages, 
FFA has difficulties in searching compact binary systems such as DNS systems. 
The orbital motion of the pulsar induces Doppler shifts that cause changes in the rotation period of pulsar. 
Thus, the sensitivity of FFA searches is reduced. 
In the frequency domain, the acceleration search techniques \citep{2018ApJ...863L..13A, 2018PhDT.......225C} partially mitigate this problem by approximating the orbital acceleration as constant over a small fraction of the orbital period. 
However, their performance deteriorates significantly when large values of the $z_{\max}$ (maximum Fourier drift parameter) are required, 
limiting their efficiency for highly accelerated systems.  

To address this challenge, 
the code \textsc{PYSOLATOR} can be used to resample the time-series data with the orbital parameters of the DNS system
and thus removes the pulsar spin period variation.
This approach restores the phase coherence of the signal and maximizes both the sensitivity and computational efficiency of the search, 
even in strongly accelerated binary systems.  
The S/N values in table \ref{tab:DNS_snr} show that after resampling the S/N were at least 30\% higher than before.
Among these signals, the periodic signal of B2127+11C can only be detected using the FFA after removing the orbital modulation effects.
These results indicate that the combined use of \textsc{PYSOLATOR} and FFA can enhance the search sensitivity of companion stars in DNS systems. Applying our pipeline to all publicly available pre-2008 archival observations of double pulsar J0737–3039A/B successfully recovers the periodic signal of pulsar A, while no detectable signal from pulsar B is present in the observational data; this is consistent with the well-known intermittency and long-term disappearance of pulsar B due to relativistic spin precession, indicating that the nondetection reported here reflects existing observational constraints rather than a limitation of the pipeline.

\begin{table*}[!ht]  
    \centering  
    \caption{Basic Information and SNR Changes of Double Neutron Star Systems}
    \label{tab:DNS_snr}
    \begin{tabular}{cccccc}
        \hline  
        Name & Obs Date & Spin & Obs & S/N & S/N \\
        & yyyy--mm--dd & Period (s) & Duration (s) & Before & After \\
        \hline  
        J0453+1559  & 2021-10-21 & 0.04578 & 10200 & 2442.9 & 5525.4 \\ 
        J0509+3801  & 2022-05-05 & 0.07654 & 5400 & 195.9 & 775.3 \\ 
        J1411+2551  & 2023-04-11 & 0.06245 & 3300 & 28.4 & 59.3 \\ 
        J1518+4904  & 2021-11-18 & 0.04093 & 10500 & 7018.8 & 9422.2 \\ 
        B1534+12    & 2023-05-21 & 0.03790 & 1800 & 68.2 & 214.2 \\ 
        J1759+5036  & 2024-12-21 & 0.17602 & 6600 & 687.8 & 973.8 \\ 
        J1829+2456  & 2023-07-25 & 0.04101 & 1800 & 82.8 & 214.2 \\ 
        J1906+0746  & 2020-09-02 & 0.14407 & 1800 & 26.6 & 140.0 \\ 
        J1913+1102  & 2022-01-20 & 0.02729 & 3600 & 45.9 & 155.4 \\ 
        B1913+16    & 2023-01-05 & 0.05903 & 3200 & 227.0 & 1589.7 \\ 
        J1946+2052  & 2024-11-01 & 0.01696 & 6600 & 38.7 & 241.5 \\ 
        B2127+11C   & 2019-11-09 & 0.03053 & 4800 & $\star$ & 7.2 \\ 
        J2150+3427  & 2022-10-31 & 0.65427 & 7335 & 130.1 & 171.7 \\ 
        \hline  
    \end{tabular}
    \begin{flushleft}  
        \textbf{Note:} 
        Column 2 lists the observation date of each double neutron star (DNS) source; 
        Column 3 lists the spin period of the primary star; 
        Column 4 lists the observation duration; 
        Columns 5 and 6 list the signal-to-noise ratios (S/N) obtained by FFA before and after resampling with \textsc{PYSOLATOR}, respectively.  
        $\star$ denotes no valid data.
    \end{flushleft}
\end{table*}

\subsection{The Narrow Radiation Beam of Companion Star}

In DNS systems, the second-formed neutron star (companion star) is usually a non-recycled \citep{2017ApJ...846..170T}, often possessing a long spin period and narrow beam opening angle.
The beam opening angle is increase as the spin period \( P \) decreases, 
and can be approximately estimated as
\begin{equation}
\rho \approx 5.4^\circ P^{-0.5}
\end{equation}
\citep{1998MNRAS.298..625T,1990nsf....8917722R}.
So, a millisecond pulsar with a spin period of 0.01 s may have a beam opening angle of $54^\circ$, 
while a pulsar with a 1 s period would have an opening angle of only about $5.4^\circ$.
This supports the observational phenomenon that most detected pulsar in DNS systems are recycled pulsars, 
while the companion star is often unseen.
In addition, for pulsars with narrow pulses (duty cycle $\delta < 1 \%$), 
more energy in the frequency domain is distributed in harmonics. 
Searching in frequency domain will loss more energy \citep{2017MNRAS.468.1994C} due to the limited number of harmonic superpositions.
Therefore, FFA is no doubt suitable for searching such periodic signals of companion stars.

It is worth noting that the detected pulsar in J1906+0746 is a young one,
indicating that its companion can be a recycled pulsar with relatively short spin period and wide beam opening angle, will be visible as a double pulsar in the future \citep{2025A&A...701A.180W}.
In addition, as in the extremely dense core-collapsing globular cluster M15, 
the companion of B2127+11C may also be a recycled one.
Thus, in FAST sky, these two targets should have higher priority in the companion signal search.

\subsection{Geodetic Precession}

Geodetic precession, predicted by general relativity, 
causes a pulsar’s spin axis (and radio beam) to gradually change direction.
The geodesic precession rates of five DNS systems are listed in Table \ref{tab:Geodetic precession}. 
This phenomenon influences whether the companion’s radio emission can be detected in DNS systems. 
For example, PSR J0737–3039B has not been visible after 2008 \citep{2012ApJ...750..130P}, 
B1913+16 is expected to disappear around 2025 \citep{2003nlgd.conf..505K}, 
and B2127+11C may become visible again between the year 2041 and 2053 \citep{2018IAUS..337..251R}.
Therefore long-term timing observations combined with precession geometry modeling can be used to predict visibility windows and identify epochs when periodic signals become detectable.

Similar to the previous subsection, pulsar J1906+0746 and B2127+11C with more than 1~$^\circ$/yr geodetic precessions.
With the $\rm \sim$8~$^\circ$/yr, it is also important to perform long term observation of J1946+2052.

\begin{table*}[!ht]
    \centering
    \caption{DNS systems with geodetic precession}
    \label{tab:Geodetic precession}
    \begin{tabular}{cc}
    \hline
        Name & Geodetic Precession \\ 
         & ($^\circ$/yr) \\
    \hline
        B1534+12$^1$  & $0.59(^{+0.12}_{-0.08})$ \\ 
        J1906+0746$^2$ & 2.17 $\pm$ 0.11 \\ 
        B1913+16$^3$  & 1.21 \\ 
        J1946+2052$^4$ & 7.96 \\ 
        B2127+11C$^5$ & 1.31 \\ 
        \hline
    \end{tabular}
    \begin{flushleft}  
    \textbf{Note:} This table lists DNS systems for which geodetic precession values have been measured. References: 1 \citep{2014ApJ...787...82F}, 2 \citep{2019Sci...365.1013D}, 3 \citep{1998ApJ...509..856K}, 4 \citep{2024ApJ...966...46M}, 5 \citep{2014A&A...565A..43K}.
    \end{flushleft}
\end{table*}

\subsection{Other newly discovered DNS Systems}

There are four newly DNS systems (or DNS candidates): 
J1953+1847D \citep{2025ApJS..279...51L} in globular cluster M71, 
J1901+0658 \citep{2024MNRAS.530.1506S}, 
J0528+3529 and J1844-0128 \citep{2025RAA....25a4003W} were discovered by FAST (the basic parameter see Table \ref{tab:new_dns}). 
The FAST globular cluster pulsar survey have enough M71 data and we will search for the companion signal soon.
For other three, no available archive data found.

\begin{table*}[!ht]
\centering
\caption{Basic parameters of four recently reported DNS systems.}
\label{tab:new_dns}
\begin{tabular}{cccccccc}
\hline
Name & DM & Spin & $P_{\rm b}$ & $\tau_{\rm age}$  & e & $M_{\rm TOT}$ & $M_{\rm c}$  \\
  & $(\text{pc cm}^{-3})$ & Period (s) & (days) & (Myr) &  & (\(M_\odot\)) & (\(M_\odot\))  \\
\hline
J1953+1847D (M71D)$^{1}$ & 117.7 & 0.10068 & 10.938 & $>8800$ & 0.628921(3) & 2.63(8) & $>1.29$  \\
J1901+0658$^{2}$ & 125.9 & 0.07574 &14.455 & $\sim 5500$ & 0.3662392(5) & 2.79(7) & $>1.11$  \\
J0528+3529$^{3}$ & 111.8 & 0.07823 & 11.726 & $\sim 1686$ & 0.2901088(10) & 2.90(12) & $>1.15$  \\
J1844-0128$^{3}$ & 138.3 & 0.02914 & 4.188 & $\sim 1500$ & 0.234949(5) & 1.7(8) & $>0.72$  \\
\hline
\end{tabular}
\begin{flushleft}
\textbf{Note:} Parameters are taken from the timing analyses in the cited papers. 
The columns represent, respectively, dispersion measure, spin period, orbital period, characteristic age, orbital eccentricity, total system mass, and companion mass. 
References: 
1 \citep{2025ApJS..279...51L}, 
2 \citep{2024MNRAS.530.1506S}, 
3 \citep{{2025RAA....25a4003W}}.
\end{flushleft}
\end{table*}

\section{Conclusion}
\label{sect:conclusion}

The conclusions of our study are as follows:

1. We introduced a FFA-based binary pulsar search pipeline, 
aiming to find the periodic signals of the companion stars in the DNS systems.
As most of these systems have a detected recycled pulsar and a highly possible normal (long spin period) pulsar, 
we suggest that FFA-based pulsar search should obtain higher sensitivity.

2. We searched most of the archived FAST data of the 13 DNS systems, with no candidate found.

3. The timing of pulsar J1946+2052 may not precious enough for orbital movement removal, 
thus we infer that maybe the FFT based or even spectra stacking  \citep[e.g.][]{2016MNRAS.459L..26P} may have chance to find its companion signal.

4. The geodetic procession may result in the next double pulsar.
The candidate can be J1906+0746 (2.17~$^\circ$/yr) and J1946+2052 (7.96~$^\circ$/yr).

5. In the future, we plan to search for the single pulses of DNS systems and determine whether the single pulses come from the companion star by comparing the pulse phase of the single pulses with that of the known pulsar.

\begin{acknowledgements}
We thank the referee's valuable suggestions, which helped to refine our paper. We are grateful to Alessandro Ridolfi for his helpful suggestions and assistance in clarifying several technical aspects of PYSOLATOR.
This work is supported by the National Key R \& D Program of China No. 2022YFC2205202, No. 2020SKA0120100, and the National Natural Science Foundation of China (NSFC, Grant Nos. 12373032, 12003047, 11773041, U2031119, 12173052, and 12173053.
Both Lei Qian and Zhichen Pan were supported by the Youth Innovation Promotion Association of CAS (id.~2018075, Y2022027, and 2023064) and the CAS "Light of West China" Program.
Liyun Zhang has been supported by the Science and Technology Program of Guizhou Province under project No.QKHPTRC-ZDSYS[2023]003 and QKHFQ[2023]003. 
This work made use of the data from FAST (Five-hundred-meter Aperture Spherical radio Telescope) (https://cstr.cn/31116.02.FAST). 
FAST is a Chinese national mega-science facility, operated by the National Astronomical Observatories, Chinese Academy of Sciences. 
We sincerely appreciate all the PIs of the
FAST archival data used in this work.

\end{acknowledgements}

\bibliographystyle{raa}
\bibliography{msbib}

@ARTICLE{2017PhRvL.119p1101A,
       author = {{Abbott}, B.~P. and {Abbott}, R. and {Abbott}, T.~D. and {Acernese}, F. and {Ackley}, K. and {Adams}, C. and {Adams}, T. and {Addesso}, P. and {Adhikari}, R.~X. and {Adya}, V.~B. and {Affeldt}, C. and {Afrough}, M. and {Agarwal}, B. and {Agathos}, M. and {Agatsuma}, K. and {Aggarwal}, N. and {Aguiar}, O.~D. and {Aiello}, L. and {Ain}, A. and {Ajith}, P. and {Allen}, B. and {Allen}, G. and {Allocca}, A. and {Altin}, P.~A. and {Amato}, A. and {Ananyeva}, A. and {Anderson}, S.~B. and {Anderson}, W.~G. and {Angelova}, S.~V. and {Antier}, S. and {Appert}, S. and {Arai}, K. and {Araya}, M.~C. and {Areeda}, J.~S. and {Arnaud}, N. and {Arun}, K.~G. and {Ascenzi}, S. and {Ashton}, G. and {Ast}, M. and {Aston}, S.~M. and {Astone}, P. and {Atallah}, D.~V. and {Aufmuth}, P. and {Aulbert}, C. and {AultONeal}, K. and {Austin}, C. and {Avila-Alvarez}, A. and {Babak}, S. and {Bacon}, P. and {Bader}, M.~K.~M. and {Bae}, S. and {Bailes}, M. and {Baker}, P.~T. and {Baldaccini}, F. and {Ballardin}, G. and {Ballmer}, S.~W. and {Banagiri}, S. and {Barayoga}, J.~C. and {Barclay}, S.~E. and {Barish}, B.~C. and {Barker}, D. and {Barkett}, K. and {Barone}, F. and {Barr}, B. and {Barsotti}, L. and {Barsuglia}, M. and {Barta}, D. and {Barthelmy}, S.~D. and {Bartlett}, J. and {Bartos}, I. and {Bassiri}, R. and {Basti}, A. and {Batch}, J.~C. and {Bawaj}, M. and {Bayley}, J.~C. and {Bazzan}, M. and {B{\'e}csy}, B. and {Beer}, C. and {Bejger}, M. and {Belahcene}, I. and {Bell}, A.~S. and {Berger}, B.~K. and {Bergmann}, G. and {Bernuzzi}, S. and {Bero}, J.~J. and {Berry}, C.~P.~L. and {Bersanetti}, D. and {Bertolini}, A. and {Betzwieser}, J. and {Bhagwat}, S. and {Bhandare}, R. and {Bilenko}, I.~A. and {Billingsley}, G. and {Billman}, C.~R. and {Birch}, J. and {Birney}, R. and {Birnholtz}, O. and {Biscans}, S. and {Biscoveanu}, S. and {Bisht}, A. and {Bitossi}, M. and {Biwer}, C. and {Bizouard}, M.~A. and {Blackburn}, J.~K. and {Blackman}, J. and {Blair}, C.~D. and {Blair}, D.~G. and {Blair}, R.~M. and {Bloemen}, S. and {Bock}, O. and {Bode}, N. and {Boer}, M. and {Bogaert}, G. and {Bohe}, A. and {Bondu}, F. and {Bonilla}, E. and {Bonnand}, R. and {Boom}, B.~A. and {Bork}, R. and {Boschi}, V. and {Bose}, S. and {Bossie}, K. and {Bouffanais}, Y. and {Bozzi}, A. and {Bradaschia}, C. and {Brady}, P.~R. and {Branchesi}, M. and {Brau}, J.~E. and {Briant}, T. and {Brillet}, A. and {Brinkmann}, M. and {Brisson}, V. and {Brockill}, P. and {Broida}, J.~E. and {Brooks}, A.~F. and {Brown}, D.~A. and {Brown}, D.~D. and {Brunett}, S. and {Buchanan}, C.~C. and {Buikema}, A. and {Bulik}, T. and {Bulten}, H.~J. and {Buonanno}, A. and {Buskulic}, D. and {Buy}, C. and {Byer}, R.~L. and {Cabero}, M. and {Cadonati}, L. and {Cagnoli}, G. and {Cahillane}, C. and {Calder{\'o}n Bustillo}, J. and {Callister}, T.~A. and {Calloni}, E. and {Camp}, J.~B. and {Canepa}, M. and {Canizares}, P. and {Cannon}, K.~C. and {Cao}, H. and {Cao}, J. and {Capano}, C.~D. and {Capocasa}, E. and {Carbognani}, F. and {Caride}, S. and {Carney}, M.~F. and {Carullo}, G. and {Casanueva Diaz}, J. and {Casentini}, C. and {Caudill}, S. and {Cavagli{\`a}}, M. and {Cavalier}, F. and {Cavalieri}, R. and {Cella}, G. and {Cepeda}, C.~B. and {Cerd{\'a}-Dur{\'a}n}, P. and {Cerretani}, G. and {Cesarini}, E. and {Chamberlin}, S.~J. and {Chan}, M. and {Chao}, S. and {Charlton}, P. and {Chase}, E. and {Chassande-Mottin}, E. and {Chatterjee}, D. and {Chatziioannou}, K. and {Cheeseboro}, B.~D. and {Chen}, H.~Y. and {Chen}, X. and {Chen}, Y. and {Cheng}, H.-P. and {Chia}, H. and {Chincarini}, A. and {Chiummo}, A. and {Chmiel}, T. and {Cho}, H.~S. and {Cho}, M. and {Chow}, J.~H. and {Christensen}, N. and {Chu}, Q. and {Chua}, A.~J.~K. and {Chua}, S.},
        title = "{GW170817: Observation of Gravitational Waves from a Binary Neutron Star Inspiral}",
      journal = {\prl},
         year = 2017,
        month = oct,
       volume = {119},
       number = {16},
          eid = {161101},
        pages = {161101},
          doi = {10.1103/PhysRevLett.119.161101},
archivePrefix = {arXiv},
       eprint = {1710.05832},
 primaryClass = {gr-qc},
       adsurl = {https://ui.adsabs.harvard.edu/abs/2017PhRvL.119p1101A}
}

@ARTICLE{2018ApJ...863L..13A,
       author = {{Andersen}, Bridget C. and {Ransom}, Scott M.},
        title = "{A Fourier Domain {\textquotedblleft}Jerk{\textquotedblright} Search for Binary Pulsars}",
      journal = {\apjl},
         year = 2018,
        month = aug,
       volume = {863},
       number = {1},
          eid = {L13},
        pages = {L13},
          doi = {10.3847/2041-8213/aad59f},
archivePrefix = {arXiv},
       eprint = {1807.07900},
 primaryClass = {astro-ph.HE},
       adsurl = {https://ui.adsabs.harvard.edu/abs/2018ApJ...863L..13A}
}

@ARTICLE{2008Sci...321..104B,
       author = {{Breton}, Rene P. and {Kaspi}, Victoria M. and {Kramer}, Michael and {McLaughlin}, Maura A. and {Lyutikov}, Maxim and {Ransom}, Scott M. and {Stairs}, Ingrid H. and {Ferdman}, Robert D. and {Camilo}, Fernando and {Possenti}, Andrea},
        title = "{Relativistic Spin Precession in the Double Pulsar}",
      journal = {Science},
         year = 2008,
        month = jul,
       volume = {321},
       number = {5885},
        pages = {104},
          doi = {10.1126/science.1159295},
archivePrefix = {arXiv},
       eprint = {0807.2644},
 primaryClass = {astro-ph},
       adsurl = {https://ui.adsabs.harvard.edu/abs/2008Sci...321..104B}
}

@ARTICLE{2003Natur.426..531B,
       author = {{Burgay}, M. and {D'Amico}, N. and {Possenti}, A. and {Manchester}, R.~N. and {Lyne}, A.~G. and {Joshi}, B.~C. and {McLaughlin}, M.~A. and {Kramer}, M. and {Sarkissian}, J.~M. and {Camilo}, F. and {Kalogera}, V. and {Kim}, C. and {Lorimer}, D.~R.},
        title = "{An increased estimate of the merger rate of double neutron stars from observations of a highly relativistic system}",
      journal = {\nat},
         year = 2003,
        month = dec,
       volume = {426},
       number = {6966},
        pages = {531-533},
          doi = {10.1038/nature02124},
archivePrefix = {arXiv},
       eprint = {astro-ph/0312071},
 primaryClass = {astro-ph},
       adsurl = {https://ui.adsabs.harvard.edu/abs/2003Natur.426..531B}
}

@PHDTHESIS{2018PhDT.......225C,
       author = {{Cameron}, Andrew David},
        title = "{Innovative pulsar searching techniques : or fantastic pulsars and how to find them}",
       school = {Rheinische Friedrich Wilhelms University of Bonn, Germany},
         year = 2018,
        month = jan,
       adsurl = {https://ui.adsabs.harvard.edu/abs/2018PhDT.......225C}
}

@ARTICLE{2017MNRAS.468.1994C,
       author = {{Cameron}, A.~D. and {Barr}, E.~D. and {Champion}, D.~J. and {Kramer}, M. and {Zhu}, W.~W.},
        title = "{An investigation of pulsar searching techniques with the fast folding algorithm}",
      journal = {\mnras},
         year = 2017,
        month = jun,
       volume = {468},
       number = {2},
        pages = {1994-2010},
          doi = {10.1093/mnras/stx589},
archivePrefix = {arXiv},
       eprint = {1703.05581},
 primaryClass = {astro-ph.IM},
       adsurl = {https://ui.adsabs.harvard.edu/abs/2017MNRAS.468.1994C}
}

@ARTICLE{1985ApJ...294L..25D,
       author = {{Dewey}, R.~J. and {Taylor}, J.~H. and {Weisberg}, J.~M. and {Stokes}, G.~H.},
        title = "{A search for low-luminosity pulsars.}",
      journal = {\apjl},
         year = 1985,
        month = jul,
       volume = {294},
        pages = {L25-L29},
          doi = {10.1086/184502},
       adsurl = {https://ui.adsabs.harvard.edu/abs/1985ApJ...294L..25D}
}

@ARTICLE{2019Sci...365.1013D,
       author = {{Desvignes}, Gregory and {Kramer}, Michael and {Lee}, Kejia and {van Leeuwen}, Joeri and {Stairs}, Ingrid and {Jessner}, Axel and {Cognard}, Isma{\"e}l and {Kasian}, Laura and {Lyne}, Andrew and {Stappers}, Ben W.},
        title = "{Radio emission from a pulsar{\textquoteright}s magnetic pole revealed by general relativity}",
      journal = {Science},
         year = 2019,
        month = sep,
       volume = {365},
       number = {6457},
        pages = {1013-1017},
          doi = {10.1126/science.aav7272},
archivePrefix = {arXiv},
       eprint = {1909.06212},
 primaryClass = {astro-ph.HE},
       adsurl = {https://ui.adsabs.harvard.edu/abs/2019Sci...365.1013D}
}

@ARTICLE{2020Natur.583..211F,
       author = {{Ferdman}, R.~D. and {Freire}, P.~C.~C. and {Perera}, B.~B.~P. and {Pol}, N. and {Camilo}, F. and {Chatterjee}, S. and {Cordes}, J.~M. and {Crawford}, F. and {Hessels}, J.~W.~T. and {Kaspi}, V.~M. and {McLaughlin}, M.~A. and {Parent}, E. and {Stairs}, I.~H. and {van Leeuwen}, J.},
        title = "{Asymmetric mass ratios for bright double neutron-star mergers}",
      journal = {\nat},
         year = 2020,
        month = jul,
       volume = {583},
       number = {7815},
        pages = {211-214},
          doi = {10.1038/s41586-020-2439-x},
archivePrefix = {arXiv},
       eprint = {2007.04175},
 primaryClass = {astro-ph.HE},
       adsurl = {https://ui.adsabs.harvard.edu/abs/2020Natur.583..211F}
}

@ARTICLE{2014ApJ...787...82F,
       author = {{Fonseca}, Emmanuel and {Stairs}, Ingrid H. and {Thorsett}, Stephen E.},
        title = "{A Comprehensive Study of Relativistic Gravity Using PSR B1534+12}",
      journal = {\apj},
         year = 2014,
        month = may,
       volume = {787},
       number = {1},
          eid = {82},
        pages = {82},
          doi = {10.1088/0004-637X/787/1/82},
archivePrefix = {arXiv},
       eprint = {1402.4836},
 primaryClass = {astro-ph.HE},
       adsurl = {https://ui.adsabs.harvard.edu/abs/2014ApJ...787...82F}
}

@ARTICLE{2021MNRAS.500.4620H,
       author = {{Haniewicz}, H.~T. and {Ferdman}, R.~D. and {Freire}, P.~C.~C. and {Champion}, D.~J. and {Bunting}, K.~A. and {Lorimer}, D.~R. and {McLaughlin}, M.~A.},
        title = "{Precise mass measurements for the double neutron star system J1829+2456}",
      journal = {\mnras},
         year = 2021,
        month = feb,
       volume = {500},
       number = {4},
        pages = {4620-4627},
          doi = {10.1093/mnras/staa3466},
archivePrefix = {arXiv},
       eprint = {2007.07565},
 primaryClass = {astro-ph.SR},
       adsurl = {https://ui.adsabs.harvard.edu/abs/2021MNRAS.500.4620H}
}

@ARTICLE{2007ApJ...670..363H,
       author = {{Hessels}, J.~W.~T. and {Ransom}, S.~M. and {Stairs}, I.~H. and {Kaspi}, V.~M. and {Freire}, P.~C.~C.},
        title = "{A 1.4 GHz Arecibo Survey for Pulsars in Globular Clusters}",
      journal = {\apj},
         year = 2007,
        month = nov,
       volume = {670},
       number = {1},
        pages = {363-378},
          doi = {10.1086/521780},
archivePrefix = {arXiv},
       eprint = {0707.1602},
 primaryClass = {astro-ph},
       adsurl = {https://ui.adsabs.harvard.edu/abs/2007ApJ...670..363H}
}

@ARTICLE{1975ApJ...195L..51H,
       author = {{Hulse}, R.~A. and {Taylor}, J.~H.},
        title = "{Discovery of a pulsar in a binary system.}",
      journal = {\apjl},
         year = 1975,
        month = jan,
       volume = {195},
        pages = {L51-L53},
          doi = {10.1086/181708},
       adsurl = {https://ui.adsabs.harvard.edu/abs/1975ApJ...195L..51H}
}

@ARTICLE{2006ApJ...644L.113J,
       author = {{Jacoby}, B.~A. and {Cameron}, P.~B. and {Jenet}, F.~A. and {Anderson}, S.~B. and {Murty}, R.~N. and {Kulkarni}, S.~R.},
        title = "{Measurement of Orbital Decay in the Double Neutron Star Binary PSR B2127+11C}",
      journal = {\apjl},
         year = 2006,
        month = jun,
       volume = {644},
       number = {2},
        pages = {L113-L116},
          doi = {10.1086/505742},
archivePrefix = {arXiv},
       eprint = {astro-ph/0605375},
 primaryClass = {astro-ph},
       adsurl = {https://ui.adsabs.harvard.edu/abs/2006ApJ...644L.113J}
}

@ARTICLE{2019SCPMA..6259502J,
       author = {{Jiang}, Peng and {Yue}, YouLing and {Gan}, HengQian and {Yao}, Rui and {Li}, Hui and {Pan}, GaoFeng and {Sun}, JingHai and {Yu}, DongJun and {Liu}, HongFei and {Tang}, NingYu and {Qian}, Lei and {Lu}, JiGuang and {Yan}, Jun and {Peng}, Bo and {Zhang}, ShuXin and {Wang}, QiMing and {Li}, Qi and {Li}, Di and {FAST Collaboration}},
        title = "{Commissioning progress of the FAST}",
      journal = {Science China Physics, Mechanics, and Astronomy},
         year = 2019,
        month = may,
       volume = {62},
       number = {5},
          eid = {959502},
        pages = {959502},
          doi = {10.1007/s11433-018-9376-1},
archivePrefix = {arXiv},
       eprint = {1903.06324},
 primaryClass = {astro-ph.IM},
       adsurl = {https://ui.adsabs.harvard.edu/abs/2019SCPMA..6259502J}
}

@INPROCEEDINGS{2015aska.confE..40K,
       author = {{Keane}, E. and {Bhattacharyya}, B. and {Kramer}, M. and {Stappers}, B. and {Keane}, E.~F. and {Bhattacharyya}, B. and {Kramer}, M. and {Stappers}, B.~W. and {Bates}, S.~D. and {Burgay}, M. and {Chatterjee}, S. and {Champion}, D.~J. and {Eatough}, R.~P. and {Hessels}, J.~W.~T. and {Janssen}, G. and {Lee}, K.~J. and {van Leeuwen}, J. and {Margueron}, J. and {Oertel}, M. and {Possenti}, A. and {Ransom}, S. and {Theureau}, G. and {Torne}, P.},
        title = "{A Cosmic Census of Radio Pulsars with the SKA}",
    booktitle = {Advancing Astrophysics with the Square Kilometre Array (AASKA14)},
         year = 2015,
        month = apr,
          eid = {40},
        pages = {40},
          doi = {10.22323/1.215.0040},
archivePrefix = {arXiv},
       eprint = {1501.00056},
 primaryClass = {astro-ph.IM},
       adsurl = {https://ui.adsabs.harvard.edu/abs/2015aska.confE..40K}
}

@ARTICLE{2014A&A...565A..43K,
       author = {{Kirsten}, Franz and {Vlemmings}, Wouter and {Freire}, Paulo and {Kramer}, Michael and {Rottmann}, Helge and {Campbell}, Robert M.},
        title = "{Precision astrometry of pulsars and other compact radio sources in the globular cluster M15}",
      journal = {\aap},
         year = 2014,
        month = may,
       volume = {565},
          eid = {A43},
        pages = {A43},
          doi = {10.1051/0004-6361/201323239},
archivePrefix = {arXiv},
       eprint = {1403.7204},
 primaryClass = {astro-ph.HE},
       adsurl = {https://ui.adsabs.harvard.edu/abs/2014A&A...565A..43K}
}

@ARTICLE{2013ApJ...778...66K,
       author = {{Kiziltan}, B{\"u}lent and {Kottas}, Athanasios and {De Yoreo}, Maria and {Thorsett}, Stephen E.},
        title = "{The Neutron Star Mass Distribution}",
      journal = {\apj},
         year = 2013,
        month = nov,
       volume = {778},
       number = {1},
          eid = {66},
        pages = {66},
          doi = {10.1088/0004-637X/778/1/66},
archivePrefix = {arXiv},
       eprint = {1011.4291},
 primaryClass = {astro-ph.GA},
       adsurl = {https://ui.adsabs.harvard.edu/abs/2013ApJ...778...66K}
}

@ARTICLE{1998ApJ...509..856K,
       author = {{Kramer}, Michael},
        title = "{Determination of the Geometry of the PSR B1913+16 System by Geodetic Precession}",
      journal = {\apj},
         year = 1998,
        month = dec,
       volume = {509},
       number = {2},
        pages = {856-860},
          doi = {10.1086/306535},
archivePrefix = {arXiv},
       eprint = {astro-ph/9808127},
 primaryClass = {astro-ph},
       adsurl = {https://ui.adsabs.harvard.edu/abs/1998ApJ...509..856K}
}

@INPROCEEDINGS{2003nlgd.conf..505K,
       author = {{Kramer}, Michael},
        title = "{Determination of the Geometry of the PSR B1913 + 16 System by Geodetic Precession}",
    booktitle = {Nonlinear Gravitodynamics: The Lense-Thirring Effect. Edited by RUFFINI REMO \& SIGISMONDI COSTANTINO. Published by World Scientific Publishing Co. Pte. Ltd},
         year = 2003,
       editor = {{Ruffini}, Remo and {Sigismondi}, Costantino},
        month = may,
        pages = {505-509},
          doi = {10.1142/9789812564818_0044},
       adsurl = {https://ui.adsabs.harvard.edu/abs/2003nlgd.conf..505K}
}

@ARTICLE{2008ARA&A..46..541K,
       author = {{Kramer}, M. and {Stairs}, I.~H.},
        title = "{The double pulsar.}",
      journal = {\araa},
         year = 2008,
        month = sep,
       volume = {46},
        pages = {541-572},
          doi = {10.1146/annurev.astro.46.060407.145247},
       adsurl = {https://ui.adsabs.harvard.edu/abs/2008ARA&A..46..541K}
}

@ARTICLE{2021PhRvX..11d1050K,
       author = {{Kramer}, M. and {Stairs}, I.~H. and {Manchester}, R.~N. and {Wex}, N. and {Deller}, A.~T. and {Coles}, W.~A. and {Ali}, M. and {Burgay}, M. and {Camilo}, F. and {Cognard}, I. and {Damour}, T. and {Desvignes}, G. and {Ferdman}, R.~D. and {Freire}, P.~C.~C. and {Grondin}, S. and {Guillemot}, L. and {Hobbs}, G.~B. and {Janssen}, G. and {Karuppusamy}, R. and {Lorimer}, D.~R. and {Lyne}, A.~G. and {McKee}, J.~W. and {McLaughlin}, M. and {M{\"u}nch}, L.~E. and {Perera}, B.~B.~P. and {Pol}, N. and {Possenti}, A. and {Sarkissian}, J. and {Stappers}, B.~W. and {Theureau}, G.},
        title = "{Strong-Field Gravity Tests with the Double Pulsar}",
      journal = {Physical Review X},
         year = 2021,
        month = oct,
       volume = {11},
       number = {4},
          eid = {041050},
        pages = {041050},
          doi = {10.1103/PhysRevX.11.041050},
archivePrefix = {arXiv},
       eprint = {2112.06795},
 primaryClass = {astro-ph.HE},
       adsurl = {https://ui.adsabs.harvard.edu/abs/2021PhRvX..11d1050K}
}

@ARTICLE{2016PhR...621..127L,
       author = {{Lattimer}, James M. and {Prakash}, Madappa},
        title = "{The equation of state of hot, dense matter and neutron stars}",
      journal = {\physrep},
         year = 2016,
        month = mar,
       volume = {621},
        pages = {127-164},
          doi = {10.1016/j.physrep.2015.12.005},
archivePrefix = {arXiv},
       eprint = {1512.07820},
 primaryClass = {astro-ph.SR},
       adsurl = {https://ui.adsabs.harvard.edu/abs/2016PhR...621..127L}
}

@ARTICLE{2025ApJ...991...38L,
       author = {{Li}, Yaowei and {Wang}, Lin and {Qian}, Lei and {Zhang}, Liyun and {Chen}, Yujie and {Yin}, Dejiang and {Li}, Baoda and {Dai}, Yinfeng and {Eatough}, Ralph P. and {Li}, Wenze and {Jiang}, Dongyue and {Zhang}, Xingnan and {Li}, Minghui and {Lian}, Yujie and {Wu}, Yuxiao and {Liu}, Tong and {Liu}, Kuo and {Pan}, Zhichen},
        title = "{Searching for Pulsars in Globular Clusters with the Fast-folding Algorithm and a New Pulsar Discovered in M13}",
      journal = {\apj},
         year = 2025,
        month = sep,
       volume = {991},
       number = {1},
          eid = {38},
        pages = {38},
          doi = {10.3847/1538-4357/add6a5},
archivePrefix = {arXiv},
       eprint = {2505.05021},
 primaryClass = {astro-ph.HE},
       adsurl = {https://ui.adsabs.harvard.edu/abs/2025ApJ...991...38L}
}

@ARTICLE{2025ApJS..279...51L,
       author = {{Lian}, Yujie and {Pan}, Zhichen and {Zhang}, Haiyan and {Cao}, Shuo and {Freire}, P.~C.~C. and {Qian}, Lei and {Eatough}, Ralph P. and {Shao}, Lijing and {Ransom}, Scott M. and {Lorimer}, Duncan R. and {Yin}, Dejiang and {Dai}, Yinfeng and {Liu}, Kuo and {Wang}, Lin and {Wang}, Yujie and {Zhang}, Zhongli and {Feng}, Zhonghua and {Li}, Baoda and {Li}, Minghui and {Liu}, Tong and {Li}, Yaowei and {Peng}, Bo and {Pan}, Yu and {Wu}, Yuxiao and {Zhang}, Liyun and {Zhang}, Xingnan and {Jiang}, Peng},
        title = "{The FAST Globular Cluster Pulsar Survey (GC FANS)}",
      journal = {\apjs},
         year = 2025,
        month = aug,
       volume = {279},
       number = {2},
          eid = {51},
        pages = {51},
          doi = {10.3847/1538-4365/ade4ba},
archivePrefix = {arXiv},
       eprint = {2506.07970},
 primaryClass = {astro-ph.HE},
       adsurl = {https://ui.adsabs.harvard.edu/abs/2025ApJS..279...51L}
}

@INPROCEEDINGS{2004HEAD....8.1101L,
       author = {{Lyne}, A.},
        title = "{The Discovery and Physics of the Double Pulsar System}",
    booktitle = {AAS/High Energy Astrophysics Division \#8},
         year = 2004,
       series = {AAS/High Energy Astrophysics Division},
       volume = {8},
        month = aug,
          eid = {11.01},
        pages = {11.01},
       adsurl = {https://ui.adsabs.harvard.edu/abs/2004HEAD....8.1101L}
}

@ARTICLE{2015ApJ...812..143M,
       author = {{Martinez}, J.~G. and {Stovall}, K. and {Freire}, P.~C.~C. and {Deneva}, J.~S. and {Jenet}, F.~A. and {McLaughlin}, M.~A. and {Bagchi}, M. and {Bates}, S.~D. and {Ridolfi}, A.},
        title = "{Pulsar J0453+1559: A Double Neutron Star System with a Large Mass Asymmetry}",
      journal = {\apj},
         year = 2015,
        month = oct,
       volume = {812},
       number = {2},
          eid = {143},
        pages = {143},
          doi = {10.1088/0004-637X/812/2/143},
archivePrefix = {arXiv},
       eprint = {1509.08805},
 primaryClass = {astro-ph.HE},
       adsurl = {https://ui.adsabs.harvard.edu/abs/2015ApJ...812..143M}
}

@ARTICLE{2017ApJ...851L..29M,
       author = {{Martinez}, J.~G. and {Stovall}, K. and {Freire}, P.~C.~C. and {Deneva}, J.~S. and {Tauris}, T.~M. and {Ridolfi}, A. and {Wex}, N. and {Jenet}, F.~A. and {McLaughlin}, M.~A. and {Bagchi}, M.},
        title = "{Pulsar J1411+2551: A Low-mass Double Neutron Star System}",
      journal = {\apjl},
         year = 2017,
        month = dec,
       volume = {851},
       number = {2},
          eid = {L29},
        pages = {L29},
          doi = {10.3847/2041-8213/aa9d87},
archivePrefix = {arXiv},
       eprint = {1711.09804},
 primaryClass = {astro-ph.HE},
       adsurl = {https://ui.adsabs.harvard.edu/abs/2017ApJ...851L..29M}
}

@ARTICLE{2024ApJ...962..167M,
       author = {{McEwen}, A.~E. and {Swiggum}, J.~K. and {Kaplan}, D.~L. and {Tan}, C.~M. and {Meyers}, B.~W. and {Fonseca}, E. and {Agazie}, G.~Y. and {Chawla}, P. and {Crowter}, K. and {DeCesar}, M.~E. and {Dolch}, T. and {Dong}, F.~A. and {Fiore}, W. and {Fonseca}, E. and {Good}, D.~C. and {Istrate}, A.~G. and {Kaspi}, V.~M. and {Kondratiev}, V.~I. and {van Leeuwen}, J. and {Levin}, L. and {Lewis}, E.~F. and {Lynch}, R.~S. and {Masui}, K.~W. and {McKee}, J.~W. and {McLaughlin}, M.~A. and {Al Noori}, H. and {Parent}, E. and {Ransom}, S.~M. and {Siemens}, X. and {Spiewak}, R. and {Stairs}, I.~H.},
        title = "{The Green Bank North Celestial Cap Survey. IX. Timing Follow-up for 128 Pulsars}",
      journal = {\apj},
         year = 2024,
        month = feb,
       volume = {962},
       number = {2},
          eid = {167},
        pages = {167},
          doi = {10.3847/1538-4357/ad11f0},
archivePrefix = {arXiv},
       eprint = {2312.07471},
 primaryClass = {astro-ph.HE},
       adsurl = {https://ui.adsabs.harvard.edu/abs/2024ApJ...962..167M}
}

@ARTICLE{2024ApJ...966...46M,
       author = {{Meng}, Lingqi and {Zhu}, Weiwei and {Kramer}, Michael and {Miao}, Xueli and {Desvignes}, Gregory and {Shao}, Lijing and {Hu}, Huanchen and {Freire}, Paulo C.~C. and {Zhang}, Yongkun and {Xue}, Mengyao and {Fang}, Ziyao and {Champion}, David J. and {Yuan}, Mao and {Miao}, Chenchen and {Niu}, Jiarui and {Fu}, Qiuyang and {Yao}, Jumei and {Guo}, Yanjun and {Zhang}, Chengmin},
        title = "{The Relativistic Spin Precession in the Compact Double Neutron Star System PSR J1946+2052}",
      journal = {\apj},
         year = 2024,
        month = may,
       volume = {966},
       number = {1},
          eid = {46},
        pages = {46},
          doi = {10.3847/1538-4357/ad381c},
archivePrefix = {arXiv},
       eprint = {2403.17828},
 primaryClass = {astro-ph.HE},
       adsurl = {https://ui.adsabs.harvard.edu/abs/2024ApJ...966...46M}
}

@ARTICLE{2025arXiv251012506M,
       author = {{Meng}, Lingqi and {Freire}, Paulo C.~C. and {Stovall}, Kevin and {Wex}, Norbert and {Miao}, Xueli and {Zhu}, Weiwei and {Kramer}, Michael and {Cordes}, James M. and {Hu}, Huanchen and {Jiang}, Jinchen and {Parent}, Emilie and {Shao}, Lijing and {Stairs}, Ingrid H. and {Xue}, Mengyao and {Brazier}, Adam and {Camilo}, Fernando and {Champion}, David J. and {Chatterjee}, Shami and {Crawford}, Fronefield and {Fang}, Ziyao and {Fu}, Qiuyang and {Guo}, Yanjun and {Hessels}, Jason W.~T. and {MacLaughlin}, Maura and {Miao}, Chenchen and {Niu}, Jiarui and {Wu}, Ziwei and {Yao}, Jumei and {Yuan}, Mao and {Yue}, Youlin and {Zhang}, Chengmin},
        title = "{The double neutron star PSR J1946+2052 I. Masses and tests of general relativity}",
      journal = {arXiv e-prints},
         year = 2025,
        month = oct,
          eid = {arXiv:2510.12506},
        pages = {arXiv:2510.12506},
          doi = {10.48550/arXiv.2510.12506},
archivePrefix = {arXiv},
       eprint = {2510.12506},
 primaryClass = {astro-ph.HE},
       adsurl = {https://ui.adsabs.harvard.edu/abs/2025arXiv251012506M}
}

@ARTICLE{2020MNRAS.497.4654M,
       author = {{Morello}, V. and {Barr}, E.~D. and {Stappers}, B.~W. and {Keane}, E.~F. and {Lyne}, A.~G.},
        title = "{Optimal periodicity searching: revisiting the fast folding algorithm for large-scale pulsar surveys}",
      journal = {\mnras},
         year = 2020,
        month = oct,
       volume = {497},
       number = {4},
        pages = {4654-4671},
          doi = {10.1093/mnras/staa2291},
archivePrefix = {arXiv},
       eprint = {2004.03701},
 primaryClass = {astro-ph.IM},
       adsurl = {https://ui.adsabs.harvard.edu/abs/2020MNRAS.497.4654M}
}

@ARTICLE{2011IJMPD..20..989N,
       author = {{Nan}, Rendong and {Li}, Di and {Jin}, Chengjin and {Wang}, Qiming and {Zhu}, Lichun and {Zhu}, Wenbai and {Zhang}, Haiyan and {Yue}, Youling and {Qian}, Lei},
        title = "{The Five-Hundred Aperture Spherical Radio Telescope (fast) Project}",
      journal = {International Journal of Modern Physics D},
         year = 2011,
        month = jan,
       volume = {20},
       number = {6},
        pages = {989-1024},
          doi = {10.1142/S0218271811019335},
archivePrefix = {arXiv},
       eprint = {1105.3794},
 primaryClass = {astro-ph.IM},
       adsurl = {https://ui.adsabs.harvard.edu/abs/2011IJMPD..20..989N}
}

@ARTICLE{2016ARA&A..54..401O,
       author = {{{\"O}zel}, Feryal and {Freire}, Paulo},
        title = "{Masses, Radii, and the Equation of State of Neutron Stars}",
      journal = {\araa},
         year = 2016,
        month = sep,
       volume = {54},
        pages = {401-440},
          doi = {10.1146/annurev-astro-081915-023322},
archivePrefix = {arXiv},
       eprint = {1603.02698},
 primaryClass = {astro-ph.HE},
       adsurl = {https://ui.adsabs.harvard.edu/abs/2016ARA&A..54..401O}
}

@ARTICLE{2016MNRAS.459L..26P,
       author = {{Pan}, Z. and {Hobbs}, G. and {Li}, D. and {Ridolfi}, A. and {Wang}, P. and {Freire}, P.},
        title = "{Discovery of two new pulsars in 47 Tucanae (NGC 104)}",
      journal = {\mnras},
         year = 2016,
        month = jun,
       volume = {459},
       number = {1},
        pages = {L26-L30},
          doi = {10.1093/mnrasl/slw037},
archivePrefix = {arXiv},
       eprint = {1603.01348},
 primaryClass = {astro-ph.HE},
       adsurl = {https://ui.adsabs.harvard.edu/abs/2016MNRAS.459L..26P}
}

@ARTICLE{2012ApJ...750..130P,
       author = {{Perera}, B.~B.~P. and {Lomiashvili}, D. and {Gourgouliatos}, K.~N. and {McLaughlin}, M.~A. and {Lyutikov}, M.},
        title = "{PSR J0737-3039B: A Probe of Radio Pulsar Emission Heights}",
      journal = {\apj},
         year = 2012,
        month = may,
       volume = {750},
       number = {2},
          eid = {130},
        pages = {130},
          doi = {10.1088/0004-637X/750/2/130},
archivePrefix = {arXiv},
       eprint = {1203.0763},
 primaryClass = {astro-ph.GA},
       adsurl = {https://ui.adsabs.harvard.edu/abs/2012ApJ...750..130P}
}

@MISC{1990nsf....8917722R,
       author = {{Rankin}, Joanna M},
        title = "{The Emission Geometry, Radiation Processes, and Evolution of Radio Pulsars}",
 howpublished = {NSF Award Number 8917722. Directorate for Mathematical and Physical Sciences, Division Of Astronomical Sciences. 1990.},
         year = 1990,
        month = feb,
        pages = {17722},
       adsurl = {https://ui.adsabs.harvard.edu/abs/1990nsf....8917722R}
}

@PHDTHESIS{2001PhDT.......123R,
       author = {{Ransom}, Scott Mitchell},
        title = "{New search techniques for binary pulsars}",
       school = {Harvard University, Massachusetts},
         year = 2001,
        month = jan,
       adsurl = {https://ui.adsabs.harvard.edu/abs/2001PhDT.......123R}
}

@ARTICLE{2002AJ....124.1788R,
       author = {{Ransom}, Scott M. and {Eikenberry}, Stephen S. and {Middleditch}, John},
        title = "{Fourier Techniques for Very Long Astrophysical Time-Series Analysis}",
      journal = {\aj},
         year = 2002,
        month = sep,
       volume = {124},
       number = {3},
        pages = {1788-1809},
          doi = {10.1086/342285},
archivePrefix = {arXiv},
       eprint = {astro-ph/0204349},
 primaryClass = {astro-ph},
       adsurl = {https://ui.adsabs.harvard.edu/abs/2002AJ....124.1788R}
}

@ARTICLE{2003ApJ...589..911R,
       author = {{Ransom}, Scott M. and {Cordes}, James M. and {Eikenberry}, Stephen S.},
        title = "{A New Search Technique for Short Orbital Period Binary Pulsars}",
      journal = {\apj},
         year = 2003,
        month = jun,
       volume = {589},
       number = {2},
        pages = {911-920},
          doi = {10.1086/374806},
archivePrefix = {arXiv},
       eprint = {astro-ph/0210010},
 primaryClass = {astro-ph},
       adsurl = {https://ui.adsabs.harvard.edu/abs/2003ApJ...589..911R}
}

@INPROCEEDINGS{2018IAUS..337..251R,
       author = {{Ridolfi}, A. and {Freire}, P.~C.~C. and {Kramer}, M. and {Bassa}, C.~G. and {Camilo}, F. and {D'Amico}, N. and {Desvignes}, G. and {Heinke}, C.~O. and {Jordan}, C. and {Lorimer}, D.~R. and {Lyne}, A. and {Manchester}, R.~N. and {Pan}, Z. and {Sarkissian}, J. and {Torne}, P. and {van den Berg}, M. and {Venkataraman}, A. and {Wex}, N.},
        title = "{Long-term observations of pulsars in the globular clusters 47 Tucanae and M15}",
    booktitle = {Pulsar Astrophysics the Next Fifty Years},
         year = 2018,
       editor = {{Weltevrede}, P. and {Perera}, B.~B.~P. and {Preston}, L.~L. and {Sanidas}, S.},
       series = {IAU Symposium},
       volume = {337},
        month = aug,
        pages = {251-254},
          doi = {10.1017/S1743921317009334},
archivePrefix = {arXiv},
       eprint = {1711.06086},
 primaryClass = {astro-ph.HE},
       adsurl = {https://ui.adsabs.harvard.edu/abs/2018IAUS..337..251R}
}

@ARTICLE{1969IEEEP..57..724S,
       author = {{Staelin}, D.~H.},
        title = "{Fast folding algorithm for detection of periodic pulse trains.}",
      journal = {IEEE Proceedings},
         year = 1969,
        month = jan,
       volume = {57},
        pages = {724-725},
          doi = {10.1109/PROC.1969.7051},
       adsurl = {https://ui.adsabs.harvard.edu/abs/1969IEEEP..57..724S}
}

@ARTICLE{2003LRR.....6....5S,
       author = {{Stairs}, Ingrid H.},
        title = "{Testing General Relativity with Pulsar Timing}",
      journal = {Living Reviews in Relativity},
         year = 2003,
        month = dec,
       volume = {6},
       number = {1},
          eid = {5},
        pages = {5},
          doi = {10.12942/lrr-2003-5},
archivePrefix = {arXiv},
       eprint = {astro-ph/0307536},
 primaryClass = {astro-ph},
       adsurl = {https://ui.adsabs.harvard.edu/abs/2003LRR.....6....5S}
}

@ARTICLE{2018ApJ...854L..22S,
       author = {{Stovall}, K. and {Freire}, P.~C.~C. and {Chatterjee}, S. and {Demorest}, P.~B. and {Lorimer}, D.~R. and {McLaughlin}, M.~A. and {Pol}, N. and {van Leeuwen}, J. and {Wharton}, R.~S. and {Allen}, B. and {Boyce}, M. and {Brazier}, A. and {Caballero}, K. and {Camilo}, F. and {Camuccio}, R. and {Cordes}, J.~M. and {Crawford}, F. and {Deneva}, J.~S. and {Ferdman}, R.~D. and {Hessels}, J.~W.~T. and {Jenet}, F.~A. and {Kaspi}, V.~M. and {Knispel}, B. and {Lazarus}, P. and {Lynch}, R. and {Parent}, E. and {Patel}, C. and {Pleunis}, Z. and {Ransom}, S.~M. and {Scholz}, P. and {Seymour}, A. and {Siemens}, X. and {Stairs}, I.~H. and {Swiggum}, J. and {Zhu}, W.~W.},
        title = "{PALFA Discovery of a Highly Relativistic Double Neutron Star Binary}",
      journal = {\apjl},
         year = 2018,
        month = feb,
       volume = {854},
       number = {2},
          eid = {L22},
        pages = {L22},
          doi = {10.3847/2041-8213/aaad06},
archivePrefix = {arXiv},
       eprint = {1802.01707},
 primaryClass = {astro-ph.HE},
       adsurl = {https://ui.adsabs.harvard.edu/abs/2018ApJ...854L..22S}
}

@ARTICLE{2024ApJ...971..150S,
       author = {{Susobhanan}, Abhimanyu and {Kaplan}, David L. and {Archibald}, Anne M. and {Luo}, Jing and {Ray}, Paul S. and {Pennucci}, Timothy T. and {Ransom}, Scott M. and {Agazie}, Gabriella and {Fiore}, William and {Larsen}, Bjorn and {O'Neill}, Patrick and {van Haasteren}, Rutger and {Anumarlapudi}, Akash and {Bachetti}, Matteo and {Bhakta}, Deven and {Champagne}, Chloe A. and {Cromartie}, H. Thankful and {Demorest}, Paul B. and {Jennings}, Ross J. and {Kerr}, Matthew and {Levina}, Sasha and {McEwen}, Alexander and {Shapiro-Albert}, Brent J. and {Swiggum}, Joseph K.},
        title = "{PINT: Maximum-likelihood Estimation of Pulsar Timing Noise Parameters}",
      journal = {\apj},
         year = 2024,
        month = aug,
       volume = {971},
       number = {2},
          eid = {150},
        pages = {150},
          doi = {10.3847/1538-4357/ad59f7},
archivePrefix = {arXiv},
       eprint = {2405.01977},
 primaryClass = {astro-ph.IM},
       adsurl = {https://ui.adsabs.harvard.edu/abs/2024ApJ...971..150S}
}

@ARTICLE{2024MNRAS.530.1506S,
       author = {{Su}, W.~Q. and {Han}, J.~L. and {Yang}, Z.~L. and {Wang}, P.~F. and {Yuan}, J.~P. and {Wang}, C. and {Zhou}, D.~J. and {Wang}, T. and {Yan}, Y. and {Jing}, W.~C. and {Cai}, N.~N. and {Xie}, L. and {Xu}, J. and {Wang}, H.~G. and {Xu}, R.~X. and {You}, X.~P.},
        title = "{The FAST Galactic Plane Pulsar Snapshot Survey - V. PSR J1901+0658 in a double neutron star system}",
      journal = {\mnras},
         year = 2024,
        month = may,
       volume = {530},
       number = {2},
        pages = {1506-1511},
          doi = {10.1093/mnras/stae888},
archivePrefix = {arXiv},
       eprint = {2403.11635},
 primaryClass = {astro-ph.HE},
       adsurl = {https://ui.adsabs.harvard.edu/abs/2024MNRAS.530.1506S}
}

@ARTICLE{2024ApJ...966...26T,
       author = {{Tan}, Chia Min and {Fonseca}, Emmanuel and {Crowter}, Kathryn and {Dong}, Fengqiu Adam and {Kaspi}, Victoria M. and {Masui}, Kiyoshi W. and {McKee}, James W. and {Meyers}, Bradley W. and {Ransom}, Scott M. and {Stairs}, Ingrid H.},
        title = "{High-cadence Timing of Binary Pulsars with CHIME}",
      journal = {\apj},
         year = 2024,
        month = may,
       volume = {966},
       number = {1},
          eid = {26},
        pages = {26},
          doi = {10.3847/1538-4357/ad28b2},
archivePrefix = {arXiv},
       eprint = {2402.08188},
 primaryClass = {astro-ph.HE},
       adsurl = {https://ui.adsabs.harvard.edu/abs/2024ApJ...966...26T}
}

@ARTICLE{2017ApJ...846..170T,
       author = {{Tauris}, T.~M. and {Kramer}, M. and {Freire}, P.~C.~C. and {Wex}, N. and {Janka}, H.-T. and {Langer}, N. and {Podsiadlowski}, Ph. and {Bozzo}, E. and {Chaty}, S. and {Kruckow}, M.~U. and {van den Heuvel}, E.~P.~J. and {Antoniadis}, J. and {Breton}, R.~P. and {Champion}, D.~J.},
        title = "{Formation of Double Neutron Star Systems}",
      journal = {\apj},
         year = 2017,
        month = sep,
       volume = {846},
       number = {2},
          eid = {170},
        pages = {170},
          doi = {10.3847/1538-4357/aa7e89},
archivePrefix = {arXiv},
       eprint = {1706.09438},
 primaryClass = {astro-ph.HE},
       adsurl = {https://ui.adsabs.harvard.edu/abs/2017ApJ...846..170T}
}

@ARTICLE{1998MNRAS.298..625T,
       author = {{Tauris}, T.~M. and {Manchester}, R.~N.},
        title = "{On the Evolution of Pulsar Beams}",
      journal = {\mnras},
         year = 1998,
        month = aug,
       volume = {298},
       number = {3},
        pages = {625-636},
          doi = {10.1046/j.1365-8711.1998.01369.x},
       adsurl = {https://ui.adsabs.harvard.edu/abs/1998MNRAS.298..625T}
}

@INCOLLECTION{2006csxs.book..623T,
       author = {{Tauris}, T.~M. and {van den Heuvel}, E.~P.~J.},
        title = "{Formation and evolution of compact stellar X-ray sources}",
    booktitle = {Compact stellar X-ray sources},
         year = 2006,
       editor = {{Lewin}, Walter H.~G. and {van der Klis}, Michiel},
       volume = {39},
        pages = {623-665},
    publisher = {Cambridge University Press},
          doi = {10.48550/arXiv.astro-ph/0303456},
       adsurl = {https://ui.adsabs.harvard.edu/abs/2006csxs.book..623T}
}

@ARTICLE{1982ApJ...253..908T,
       author = {{Taylor}, J.~H. and {Weisberg}, J.~M.},
        title = "{A new test of general relativity - Gravitational radiation and the binary pulsar PSR 1913+16}",
      journal = {\apj},
         year = 1982,
        month = feb,
       volume = {253},
        pages = {908-920},
          doi = {10.1086/159690},
       adsurl = {https://ui.adsabs.harvard.edu/abs/1982ApJ...253..908T}
}

@ARTICLE{2015ApJ...798..118V,
       author = {{van Leeuwen}, J. and {Kasian}, L. and {Stairs}, I.~H. and {Lorimer}, D.~R. and {Camilo}, F. and {Chatterjee}, S. and {Cognard}, I. and {Desvignes}, G. and {Freire}, P.~C.~C. and {Janssen}, G.~H. and {Kramer}, M. and {Lyne}, A.~G. and {Nice}, D.~J. and {Ransom}, S.~M. and {Stappers}, B.~W. and {Weisberg}, J.~M.},
        title = "{The Binary Companion of Young, Relativistic Pulsar J1906+0746}",
      journal = {\apj},
         year = 2015,
        month = jan,
       volume = {798},
       number = {2},
          eid = {118},
        pages = {118},
          doi = {10.1088/0004-637X/798/2/118},
archivePrefix = {arXiv},
       eprint = {1411.1518},
 primaryClass = {astro-ph.SR},
       adsurl = {https://ui.adsabs.harvard.edu/abs/2015ApJ...798..118V}
}

@ARTICLE{2016ApJ...829...55W,
       author = {{Weisberg}, J.~M. and {Huang}, Y.},
        title = "{Relativistic Measurements from Timing the Binary Pulsar PSR B1913+16}",
      journal = {\apj},
         year = 2016,
        month = sep,
       volume = {829},
       number = {1},
          eid = {55},
        pages = {55},
          doi = {10.3847/0004-637X/829/1/55},
archivePrefix = {arXiv},
       eprint = {1606.02744},
 primaryClass = {astro-ph.HE},
       adsurl = {https://ui.adsabs.harvard.edu/abs/2016ApJ...829...55W}
}

@software{2020ascl.soft03012R,
       author = {{Ridolfi}, Alessandro},
        title = "{PYSOLATOR: Remove orbital modulation from a binary pulsar and/or its companion}",
 howpublished = {Astrophysics Source Code Library, record ascl:2003.012},
         year = 2020,
        month = mar,
          eid = {ascl:2003.012},
archivePrefix = {ascl},
       eprint = {2003.012},
       adsurl = {https://ui.adsabs.harvard.edu/abs/2020ascl.soft03012R}
}

\end{document}